# Perturbative Renormalization of Lattice Bilinear Quark Operators

M. Göckeler[1,2], R. Horsley[3], E.-M. Ilgenfritz[3], H. Perlt[4], P. Rakow[1],
G. Schierholz[5,1] and A. Schiller[4]

[1] Gruppe Theorie der Elementarteilchen,
Höchstleistungsrechenzentrum HLRZ, c/o Forschungszentrum Jülich,
D-52425 Jülich, Germany

[2] Institut für Theoretische Physik, RWTH Aachen,
D-52056 Aachen, Germany

[3] Institut für Physik, Humboldt-Universität,
D-10115 Berlin, Germany

[4] Institut für Theoretische Physik, Universität Leipzig,
D-04109 Leipzig, Germany

[5] Deutsches Elektronen-Synchrotron DESY,
D-22603 Hamburg, Germany

**Abstract**

Our aim is to compute the lower moments of the unpolarized and polarized deep-inelastic structure functions of the nucleon on the lattice. The theoretical basis of the calculation is the operator product expansion. To construct operators with the appropriate continuum behavior out of the bare lattice operators one must absorb the effects of momentum scales far greater than any physical scale into a renormalization of the operators. In this work we compute the renormalization constants of all bilinear quark operators of leading twist and spin up to four. The calculation is done for Wilson fermions and in the quenched approximation where dynamical quark loops are neglected.

# 1 Introduction

This work is part of an ongoing effort [1, 2] to compute the deep-inelastic structure functions of the nucleon both for unpolarized and polarized beams and targets. The theoretical basis of the calculation is the operator product expansion (OPE). The OPE relates the moments of the structure functions to forward nucleon matrix elements of local operators. Lattice simulations of these matrix elements, combined with a perturbative evaluation of the short-distance parameters of the expansion, i.e. the Wilson coefficients, can provide complete information of the deep-inelastic structure of the nucleon.

A necessary ingredient of the calculation is the renormalization of the lattice operators. The bare lattice operators are in general ultra-violet divergent. They must be renormalized such that in the limit of zero lattice spacing the renormalized operators correspond to finite, Lorentz covariant operators which obey the same renormalization conditions as those in the continuum. In particular, operator matrix elements and Wilson coefficients must be computed in the same renormalization scheme.

In principle the renormalization constants can be computed in perturbation theory, like the Wilson coefficients. However, the lattice operators may be subjected to large tadpole-induced renormalizations which lead to poor convergence of the perturbative series, and the bare coupling constant may not be the most effective expansion parameter to use [3]. Both problems can be remedied by simple redefinitions of the basic operators used to define the lattice theory and the expansion parameter, once the perturbative result is known. The validity of this procedure can finally be checked by computing the renormalization constants non-perturbatively [4, 5].

In this paper we study the renormalization of quark bilinear operators in perturbation theory to one loop order. The renormalization of gluonic operators, which enter the OPE to order $\alpha_s^2$, will be dealt with in a future publication. We shall focus on operators of leading twist and spin less or equal to four. We use Wilson fermions and work in the quenched approximation where one neglects the effects of dynamical quark loops. A brief account of our results was given in ref. [2].

The paper is organized as follows. In section 2 we present the operators and discuss their transformation properties under the hypercubic group. In section 3 we state the renormalization conditions. Section 4 contains a brief description of the calculation. The results are presented in section 5 for a few representations which are particularly suited for numerical simulations. Finally, in the appendix we give results for the full tensors, from which the renormalization constants for all other representations can be deduced.

# 2  Operators and Representations

The cross section for deep-inelastic lepton-nucleon scattering can be written in terms of four structure functions, $F_1$, $F_2$, $g_1$ and $g_2$. Here $F_1, F_2$ are the well-known spin averaged structure functions and $g_1, g_2$ are the polarized structure functions.

**OPE**

The OPE reads

$$\begin{aligned}
2\int_0^1 dx\, x^{n-1} F_1(x, Q^2) &= \sum_f c_{1,n}^{(f)}(\mu^2/Q^2, g(\mu))\, v_n^{(f)}(\mu), \\
\int_0^1 dx\, x^{n-2} F_2(x, Q^2) &= \sum_f c_{2,n}^{(f)}(\mu^2/Q^2, g(\mu))\, v_n^{(f)}(\mu), \\
2\int_0^1 dx\, x^n g_1(x, Q^2) &= \frac{1}{2}\sum_f e_{1,n}^{(f)}(\mu^2/Q^2, g(\mu))\, a_n^{(f)}(\mu),\ n=0,2,\ldots, \\
2\int_0^1 dx\, x^n g_2(x, Q^2) &= \frac{1}{2}\frac{n}{n+1}\sum_f [e_{2,n}^{(f)}(\mu^2/Q^2, g(\mu))\, d_n^{(f)}(\mu) \\
&\quad - e_{1,n}^{(f)}(\mu^2/Q^2, g(\mu))\, a_n^{(f)}(\mu)],\ n=2,4,\ldots
\end{aligned} \qquad (1)$$

with

$$\frac{1}{2}\sum_{\vec{s}} \langle \vec{p},\vec{s}|\mathcal{O}_{\{\mu_1\cdots\mu_n\}}^{(f)}|\vec{p},\vec{s}\rangle = 2v_n^{(f)}(\mu)[p_{\mu_1}\cdots p_{\mu_n} - \text{traces}], \qquad (2)$$

$$\langle \vec{p},\vec{s}|\mathcal{O}_{\{\sigma\mu_1\cdots\mu_n\}}^{5(f)}|\vec{p},\vec{s}\rangle = \frac{1}{n+1}a_n^{(f)}(\mu)[s_\sigma p_{\mu_1}\cdots p_{\mu_n} + \cdots - \text{traces}], \qquad (3)$$

$$\langle \vec{p},\vec{s}|\mathcal{O}_{[\sigma\{\mu_1]\cdots\mu_n\}}^{5(f)}|\vec{p},\vec{s}\rangle = \frac{1}{n+1}d_n^{(f)}(\mu)[(s_\sigma p_{\mu_1} - s_{\mu_1} p_\sigma)p_{\mu_2}\cdots p_{\mu_n} + \cdots - \text{traces}], \qquad (4)$$

where the superscript $f$ distinguishes the various operators and its Wilson coefficients, and $\mu$ denotes the subtraction point. In eqs. (2) – (4) $\{\cdots\}$ ($[\cdots]$) means symmetrization (anti-symmetrization) of the indices. For the definition of $\mathcal{O}_{[\sigma\{\mu_1]\cdots\mu_n\}}^{5(f)}$ see ref. [6]. In the following we shall restrict ourselves to the leading quark bilinear operators

$$\mathcal{O}_{\mu_1\cdots\mu_n} = \left(\frac{i}{2}\right)^{n-1} \bar{\psi}\gamma_{\mu_1}\overset{\leftrightarrow}{D}_{\mu_2}\cdots\overset{\leftrightarrow}{D}_{\mu_n}\psi - \text{traces}, \qquad (5)$$

$$\mathcal{O}_{\sigma\mu_1\cdots\mu_n}^5 = \left(\frac{i}{2}\right)^n \bar{\psi}\gamma_\sigma\gamma_5 \overset{\leftrightarrow}{D}_{\mu_1}\cdots\overset{\leftrightarrow}{D}_{\mu_n}\psi - \text{traces}. \qquad (6)$$

We may do so because we will be working in the quenched approximation. The corresponding gluon operators are

$$\mathcal{O}_{\mu_1\cdots\mu_n} = i^{n-2}\text{Tr}F_{\mu_1}^\alpha D_{\mu_2}\cdots D_{\mu_{n-1}}F_{\alpha\mu_n} - \text{traces}, \qquad (7)$$

$$\mathcal{O}_{\sigma\mu_1\cdots\mu_n}^5 = i^{n-1}\text{Tr}\widetilde{F}_\sigma^\alpha D_{\mu_1}\cdots D_{\mu_{n-1}}F_{\alpha\mu_n} - \text{traces}, \qquad (8)$$

where $\widetilde{F}_{\mu\nu} = \frac{1}{2}\epsilon_{\mu\nu\rho\sigma}F^{\rho\sigma}$. We will deal with the gluon operators in a future publication.

The traceless and symmetric operators $\mathcal{O}_{\{\mu_1\cdots\mu_n\}}$ and $\mathcal{O}^5_{\{\sigma\mu_1\cdots\mu_n\}}$ transform irreducibly under the Lorentz group. The r.h.s. of eqs. (2) and (3) are the only traceless, symmetric tensors of maximum spin, $n$ and $n+1$, respectively, one can build from a single momentum vector and the polarization vector $s_\mu$. Both operators have twist two. The operator $\mathcal{O}^5_{[\sigma\{\mu_1]\cdots\mu_n\}}$, which is also traceless but of mixed symmetry, transforms irreducibly as well. Again it is only possible to form a single traceless, mixed symmetric tensor from the momentum and polarization vectors, as given by the r.h.s. of eq. (4). This operator has spin $n$ and twist three.

The moments (1) are independent of $\mu$. The operator matrix elements inherit the $\mu$-dependence from the renormalization constants, which we will introduce in the next section. This means that the $\mu$-dependence of the renormalization constants must 'match' the $\mu$-dependence of the Wilson coefficients $c^{(f)}_{k,n}(\mu^2/Q^2, g(\mu))$, $e^{(f)}_{k,n}(\mu^2/Q^2, g(\mu))$ which are known perturbatively.

## Lattice Operators

We do a Wick rotation to obtain the operators in euclidean space-time. On the lattice we furthermore have to replace the covariant derivatives by the lattice covariant derivative

$$\vec{D}_\mu \psi(x) = \frac{1}{2a}\left(U_{x,\mu}\psi(x+\hat{\mu}) - U^\dagger_{x-\hat{\mu},\mu}\psi(x-\hat{\mu})\right) \tag{9}$$

$$\bar{\psi}(x)\overleftarrow{D}_\mu = \frac{1}{2a}\left(\bar{\psi}(x+\hat{\mu})U^\dagger_{x,\mu} - \bar{\psi}(x-\hat{\mu})U_{x-\hat{\mu},\mu}\right), \tag{10}$$

where $a$ is the lattice spacing. The link matrix $U_{x,\mu}$ is given by

$$U_{x,\mu} = \exp\{iagA_\mu(x)\},\ A_\mu \equiv T^a A^a_\mu, \tag{11}$$

where $g$ is the bare gauge coupling and $T^a$ are the generators of the $SU(3)$ Lie algebra.

We use the Fourier decomposition of the fermion fields and the gauge potentials

$$\psi(x) = \int \frac{d^4p}{(2\pi)^4}\psi(p)e^{ipx}, \tag{12}$$

$$\bar{\psi}(x) = \int \frac{d^4p}{(2\pi)^4}\bar{\psi}(p)e^{-ipx}, \tag{13}$$

$$A_\mu(x) = \int \frac{d^4p}{(2\pi)^4}A_\mu(p)e^{ip(x+a\hat{\mu}/2)}, \tag{14}$$

where the momenta are restricted to the first Brillouin zone. The phase factor $e^{iap_\mu/2}$ in eq. (14) amounts to redefining the potentials at the midpoints of the links connecting two neighboring lattice sites. In momentum space the operators then take the following form:

Order $g^0$

$$\mathcal{O}_{\mu_1\cdots\mu_n} = \bar{\psi}(p)\gamma_{\mu_1}\psi(p)\prod_{i=2}^{n}\frac{\sin(ap_{\mu_i})}{a} - \text{traces};\tag{15}$$

Order $g^1$

$$\begin{aligned}\mathcal{O}_{\mu_1\cdots\mu_n} &= g\bar{\psi}(p')\gamma_{\mu_1}\psi(p)\sum_{i=2}^{n}\prod_{j=2}^{i-1}\prod_{k=i+1}^{n}\frac{\sin(ap'_{\mu_j})}{a}\frac{\sin(ap_{\mu_k})}{a}\\ &\quad \times A_{\mu_i}(q)\cos(a(p'_{\mu_i}+p_{\mu_i})/2) - \text{traces},\end{aligned}\tag{16}$$

where $p' = p + q$;

Order $g^2$

$$\begin{aligned}\mathcal{O}_{\mu_1\cdots\mu_n} &= g^2\bar{\psi}(p')\gamma_{\mu_1}\psi(p)\bigg\{\sum_{i=2}^{n-1}\sum_{j=i+1}^{n}\prod_{k=2}^{i-1}\prod_{l=i+1}^{j-1}\prod_{m=j+1}^{n}\frac{\sin(ap'_{\mu_k})}{a}\\ &\quad \times \frac{\sin(a(p_{\mu_l}+q_{\mu_l}))}{a}\frac{\sin(ap_{\mu_i})}{a}\\ &\quad \times A_{\mu_i}(q')\cos(a(p'_{\mu_i}-q'_{\mu_i}/2))\,A_{\mu_j}(q)\cos(a(p_{\mu_j}+q_{\mu_j}/2))\\ &\quad -\frac{a}{2}\sum_{i=2}^{n}\prod_{j=2}^{i-1}\prod_{k=i+1}^{n}\frac{\sin(ap_{\mu_j})}{a}\frac{\sin(ap_{\mu_k})}{a}\\ &\quad \times A_{\mu_i}(q')A_{\mu_i}(q)\sin(a(p'_{\mu_i}+p_{\mu_i})/2)\bigg\} - \text{traces},\end{aligned}\tag{17}$$

where $p' = p + q + q'$.

The operators $\mathcal{O}^5_{\sigma\mu_1\cdots\mu_n}$ are obtained from eqs. (15) – (17) by replacing $\bar{\psi}(p')\gamma_{\mu_1}\psi(p)$ by $\bar{\psi}(p')\gamma_\sigma\gamma_5\psi(p)$ and by adjusting the rest of the indices appropriately.

These expressions for the operators, together with the well-known lattice Feynman rules [7, 8], will form the basis of our calculation.

### Representations

In euclidean space-time the Lorentz group is replaced by the orthogonal group $O(4)$, which on the lattice reduces to the hypercubic group $H(4) \subset O(4)$. Accordingly, the lattice operators are classified by their transformation properties under the hypercubic group and, of course, charge conjugation.

In the continuum it was relatively easy to identify the operators of leading spin transforming irreducibly under the Lorentz group. The corresponding lattice operators will in general not transform irreducibly, which allows them to mix with lower-dimensional operators under renormalization. A necessary condition for a lattice operator to be multiplicatively renormalizable is that it belongs to an irreducible representation of $H(4)$.

Under charge conjugation the operators transform as

$$\mathcal{O}_{\mu_1\mu_2\cdots\mu_n} \rightarrow (-1)^n \mathcal{O}_{\mu_1\mu_n\mu_{n-1}\cdots\mu_2}, \tag{18}$$

$$\mathcal{O}^5_{\sigma\mu_1\cdots\mu_n} \rightarrow (-1)^n \mathcal{O}^5_{\sigma\mu_n\mu_{n-1}\cdots\mu_1}. \tag{19}$$

Usually only $C = +$ operators contribute to the OPE (1). In the quenched approximation, however, also operators $\mathcal{O}_{\mu_1\cdots\mu_n}$ with odd $n$ are relevant.

In ref. [9] we have identified all irreducible representations carried by the operators $\mathcal{O}$ and $\mathcal{O}^5$ up to rank four. Furthermore we have given explicit bases for all irreducible subspaces. We will summarize our results here.

We follow the notation of ref. [10] (see also ref. [11]) and denote the irreducible representations of $H(4)$ by $\tau_k^{(l)}$, where $l$ is the dimension of the representation and $k$ distinguishes inequivalent representations of the same dimension. The defining representation will be called $\tau_1^{(4)}$. Let us first consider the operators $\mathcal{O}_{\mu_1\cdots\mu_n}$, which transform as the tensor product

$$\bigotimes_1^n \tau_1^{(4)}. \tag{20}$$

For the traceless tensors of $n = 2, 3$ and $4$ we obtain the decomposition

$$\tau_1^{(4)} \otimes \tau_1^{(4)} \stackrel{C\equiv+}{=} \underbrace{\tau_1^{(3)} \oplus \tau_3^{(6)}}_{\text{symmetric}} \oplus \tau_1^{(6)}, \tag{21}$$

$$\tau_1^{(4)} \otimes \tau_1^{(4)} \otimes \tau_1^{(4)} \stackrel{C\equiv-}{=} \underbrace{\tau_1^{(4)} \oplus \tau_2^{(4)} \oplus \tau_1^{(8)}}_{\text{symmetric}} \oplus \tau_1^{(4)} \oplus \tau_1^{(8)} \oplus \tau_2^{(8)}, \tag{22}$$

$$\tau_1^{(4)} \otimes \tau_1^{(4)} \otimes \tau_1^{(4)} \otimes \tau_1^{(4)} \stackrel{C\equiv+}{=} \underbrace{\tau_1^{(1)} \oplus \tau_2^{(1)} \oplus \tau_1^{(2)} \oplus \tau_1^{(3)} \oplus \tau_1^{(6)} \oplus \tau_2^{(6)} \oplus \tau_3^{(6)}}_{\text{symmetric}} \oplus \tau_1^{(2)} \oplus \tau_2^{(2)} \tag{23}$$
$$\oplus 2\tau_2^{(3)} \oplus 2\tau_3^{(3)} \oplus \tau_4^{(3)} \oplus 2\tau_1^{(6)} \oplus 3\tau_2^{(6)} \oplus 2\tau_3^{(6)} \oplus 3\tau_4^{(6)}.$$

We have distinguished between representations on totally symmetric tensors and tensors of different symmetry. We see that, unlike in the continuum, the representations are not uniquely determined by requiring, e.g., that the tensors are traceless and symmetric. There are several representations which appear twice, e.g., $\tau_1^{(8)}$ in eq. (22), and we will see that they can mix due to the presence of non-O(4) covariant contributions, which complicates the matter considerably.

Let us now consider the operators $\mathcal{O}^5_{\sigma\mu_1\cdots\mu_n}$ and their representations. Here one factor of $\tau_1^{(4)}$ gets replaced by $\tau_1^{(4)} \otimes \tau_4^{(1)} = \tau_4^{(4)}$. Thus the operators transform as the tensor

product

$$\Big(\bigotimes_{1}^{n+1} \tau_1^{(4)}\Big) \otimes \tau_4^{(1)}. \tag{24}$$

We are interested in the cases $n = 0$ and 2. For the traceless tensors we find the decomposition

$$\tau_1^{(4)} \otimes \tau_4^{(1)} \stackrel{C}{\equiv}{}^+ \tau_4^{(4)} \tag{25}$$

$$\tau_1^{(4)} \otimes \tau_1^{(4)} \otimes \tau_1^{(4)} \otimes \tau_4^{(1)} \stackrel{C}{\equiv}{}^+ \underbrace{\tau_3^{(4)} \oplus \tau_4^{(4)} \oplus \tau_2^{(8)}}_{\text{symmetric}} \oplus \tau_1^{(8)} \oplus \tau_2^{(8)}. \tag{26}$$

Again, there is the possibility of mixing under renormalization. Unlike in the continuum, operators of the form (5) may also mix with operators of the form (6).

The numerical simulations are greatly facilitated if one keeps the spatial momenta as small as possible. We shall therefore concentrate on those representations which require a non-zero external momentum in at most one spatial direction. The renormalization constants of the operators in the other representations will be given in the appendix.

We assume that the spatial momentum is either zero, or non-zero only in the 1-direction. Following ref. [9], we are then led to consider the operators and representations listed in table 1. It will turn out that it is not sufficient to consider the totally symmetric operator

$$\mathcal{O}_{\{114\}} - \frac{1}{2}(\mathcal{O}_{\{224\}} + \mathcal{O}_{\{334\}}) \tag{27}$$

because it mixes with the mixed symmetric operator

$$\mathcal{O}_{\langle\langle 411\rangle\rangle} - \frac{1}{2}(\mathcal{O}_{\langle\langle 422\rangle\rangle} + \mathcal{O}_{\langle\langle 433\rangle\rangle}), \tag{28}$$

as indicated by the curly brackets, where

$$\mathcal{O}_{\langle\langle \mu\nu\rho\rangle\rangle} = \mathcal{O}_{\mu\nu\rho} + \mathcal{O}_{\mu\rho\nu} - \mathcal{O}_{\rho\mu\nu} - \mathcal{O}_{\rho\nu\mu}. \tag{29}$$

Both operators belong to the representation $\tau_1^{(8)}$. In case of the operator $\mathcal{O}_{\{\mu\nu\}}$ we will consider two different representations. In the continuum limit both of them must lead to the same final result. The purpose is to check for finite cut-off effects in combination with our numerical simulations. In the last column of the table we have listed the matrix elements to which the operators contribute. Note that we have changed our notation. The operator (28) is the negative of the operator considered previously [2].

## 3   Renormalization Conditions

Let us denote the lattice regularized operators by $\mathcal{O}(a)$. We then define finite operators $\mathcal{O}(\mu)$, renormalized at the scale $\mu$, by

$$\mathcal{O}^l(\mu) = Z_{\mathcal{O}}^{lm}((a\mu)^2, g(a))\mathcal{O}^m(a), \tag{30}$$

| Rank | Operator | Representation | $C$ | $\langle \mathcal{O} \rangle$ |
|---|---|---|---|---|
| 2 | $\mathcal{O}_{\{14\}}$ | $\tau_3^{(6)}$ | $+$ | $v_2$ |
| 2 | $\mathcal{O}_{\{44\}} - \frac{1}{3}(\mathcal{O}_{\{11\}} + \mathcal{O}_{\{22\}} + \mathcal{O}_{\{33\}})$ | $\tau_1^{(3)}$ | $+$ | $v_2$ |
| 3 | $\left\{ \begin{array}{l} \mathcal{O}_{\{114\}} - \frac{1}{2}(\mathcal{O}_{\{224\}} + \mathcal{O}_{\{334\}}) \\ \mathcal{O}_{\langle\langle 411\rangle\rangle} - \frac{1}{2}(\mathcal{O}_{\langle\langle 422\rangle\rangle} + \mathcal{O}_{\langle\langle 433\rangle\rangle}) \end{array} \right\}$ | $\tau_1^{(8)}$ | $-$ | $v_3$ |
| 4 | $\mathcal{O}_{\{1144\}} + \mathcal{O}_{\{2233\}} - \mathcal{O}_{\{1133\}} - \mathcal{O}_{\{2244\}}$ | $\tau_1^{(2)}$ | $+$ | $v_4$ |
| 1 | $\mathcal{O}_2^5$ | $\tau_4^{(4)}$ | $+$ | $a_0$ |
| 3 | $\mathcal{O}_{\{214\}}^5$ | $\tau_3^{(4)}$ | $+$ | $a_2$ |
| 3 | $\mathcal{O}_{[2\{1]4\}}^5$ | $\tau_1^{(8)}$ | $+$ | $d_2$ |

Table 1: The operators and their representations.

where the superscripts $l, m$ mark different operators transforming according to the same representation.

We will assume for the moment that there is only one representation of its kind. Let us then regard the full theory with quark and gluon operators first, before we go to the quenched approximation. In this case eq. (30) becomes

$$\begin{pmatrix} \mathcal{O}^q(\mu) \\ \mathcal{O}^g(\mu) \end{pmatrix} = \begin{pmatrix} Z^{qq} & Z^{qg} \\ Z^{gq} & Z^{gg} \end{pmatrix} \begin{pmatrix} \mathcal{O}^q(a) \\ \mathcal{O}^g(a) \end{pmatrix}, \quad (31)$$

where $q$ ($g$) stands for quark (gluon). In perturbation theory one considers Green functions of the operators (5) – (8) computed between external off-shell quark and gluon states in a fixed gauge. We will impose the renormalization conditions

$$\langle q(p)|\mathcal{O}^q(\mu)|q(p)\rangle = \langle q(p)|\mathcal{O}^q(a)|q(p)\rangle \Big|_{p^2=\mu^2}^{tree}, \quad (32)$$

$$\langle q(p)|\mathcal{O}^g(\mu)|q(p)\rangle = 0, \quad (33)$$

$$\langle g(p)|\mathcal{O}^q(\mu)|g(p)\rangle = 0, \quad (34)$$

$$\langle g(p)|\mathcal{O}^g(\mu)|g(p)\rangle = \langle g(p)|\mathcal{O}^g(a)|g(p)\rangle \Big|_{p^2=\mu^2}^{tree}, \quad (35)$$

with $|q(p)\rangle$ ($|g(p)\rangle$) being a quark (gluon) state of momentum $p$. The understanding of eqs. (32) and (35) is that the conditions hold for those contributions which are proportional to the tree-level expressions. In the limit $a \to 0$ these conditions amount to the continuum, momentum subtraction renormalization scheme. One obtains operators that are multiplicatively renormalizable by diagonalizing the anomalous part of the matrix of $Z$'s in eq. (31).

Let us now consider the quenched approximation. In this approximation

$$\langle g(p)|\mathcal{O}^q(a)|g(p)\rangle = 0. \tag{36}$$

This implies furthermore that $Z^{qg} = 0$. It then follows that for the quark operators we have

$$\mathcal{O}^q(\mu) = Z^{qq}\mathcal{O}^q(a) \tag{37}$$

with the renormalization condition

$$\langle q(p)|\mathcal{O}^q(\mu)|q(p)\rangle \equiv \langle q(p)|\mathcal{O}^q(a)|q(p)\rangle\Big|^{tree}_{p^2=\mu^2} = Z^{qq}\langle q(p)|\mathcal{O}^q(a)|q(p)\rangle. \tag{38}$$

Thus, in the absence of dynamical quark loops, the (singlet) quark operators do not mix with gluon operators under renormalization.

## 4 Outline of the Calculation

The task is now to compute $\langle q(p)|\mathcal{O}(a)|q(p)\rangle$ for quark operators such as listed in table 1.

### General Considerations

We use the standard Wilson action with $r = 1$. The calculation is done for zero renormalized quark mass. We work in the Feynman gauge. The Feynman diagrams we have to calculate are listed in fig. 1. Only the operators which contain at least one covariant derivative receive contributions from the operator tadpole (fig. 1b) and cockscomb (figs. 1c,d) diagrams.

The calculation basically amounts to the computation of integrals of the form

$$\mathcal{I}_{\mu_1\cdots\mu_n}(a,p) = \int \frac{\mathrm{d}^4k}{(2\pi)^4}\mathcal{K}_{\mu_1\cdots\mu_n}(a,p,k), \tag{39}$$

where the integration is over the first Brillouin zone $-\pi/a \leq k_\mu < \pi/a$. We follow Kawai et al. [7] and write

$$\mathcal{I} = \widetilde{\mathcal{I}} + (\mathcal{I} - \widetilde{\mathcal{I}}), \tag{40}$$

where

$$\widetilde{\mathcal{I}}(a,p) = \mathcal{I}(a,0) + \sum_\alpha p_\alpha \frac{\partial}{\partial p_\alpha}\mathcal{I}(a,p)\Big|_{p=0} + \frac{1}{2!}\sum_{\alpha,\beta} p_\alpha p_\beta \frac{\partial^2}{\partial p_\alpha \partial p_\beta}\mathcal{I}(a,p)\Big|_{p=0}$$

$$+ \frac{1}{3!}\sum_{\alpha,\beta,\gamma} p_\alpha p_\beta p_\gamma \frac{\partial^3}{\partial p_\alpha \partial p_\beta \partial p_\gamma}\mathcal{I}(a,p)\Big|_{p=0} + \cdots \tag{41}$$

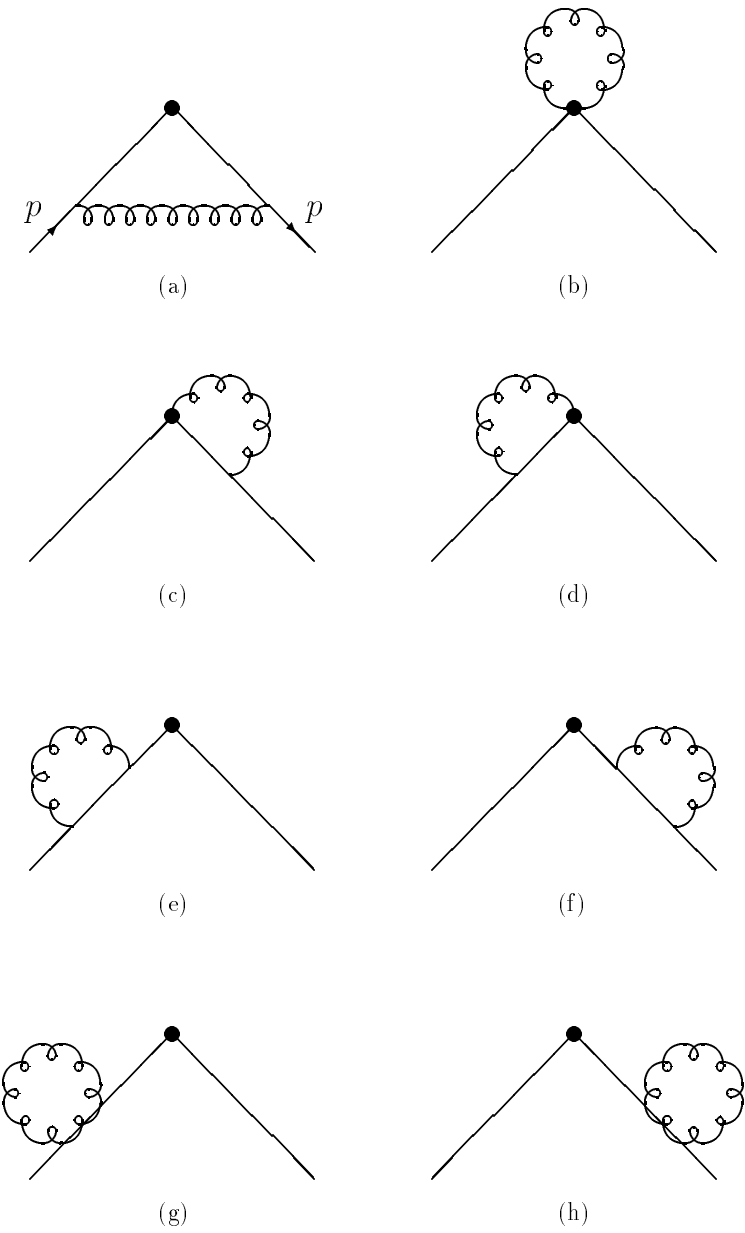

Figure 1: The lattice Feynman diagrams: (a) the vertex diagram, (b) the operator tadpole diagram, (c,d) the cockscomb diagrams, (e,f) the self-energy diagrams and (g,h) the leg tadpole diagrams.

is the Taylor expansion of the integral in the external momentum $p$. The order of the expansion is determined by the degree of ultraviolet divergence of $\mathcal{I}$, so that the integral $\mathcal{I} - \widetilde{\mathcal{I}}$ is rendered ultraviolet finite. The latter can then be computed in the continuum, i.e. taking $a \to 0$. The lattice integrals that remain to be computed do not depend on $p$, which greatly simplifies the calculation. However, the Taylor expansion about $p = 0$ will in general create an infrared divergence. In order to regularize this divergence we use dimensional regularization with $d > 4$. The infrared singularities manifest themselves as poles in $\epsilon = 4 - d$. The infrared poles of $\widetilde{\mathcal{I}}$ must cancel with those of $\mathcal{I} - \widetilde{\mathcal{I}}$. In dimensional regularization $\widetilde{\mathcal{I}}$ vanishes in the continuum limit. This can happen because the infrared divergence of $\widetilde{\mathcal{I}}$ cancels exactly its ultraviolet divergence, thus turning the ultraviolet divergent integral $\mathcal{I}$ into an ultraviolet finite, but infrared divergent expression. The ultraviolet divergent contributions of the lattice integrals will cancel in the operators which we are interested in.

The calculation is done analytically as far as this is possible. Using *Mathematica* [12], we have developed a program package which, as input, only requires to state the Feynman diagram one wants to compute in symbolic form by specifying the operator, propagators and vertices defining the loop. The program then collects the various building blocks, works out the Dirac structure and does the necessary expansions. The continuum integrals are calculated analytically, including the infrared pole terms. The lattice integrals are separated into infrared singular and infrared finite integrals. The infrared singular integrals are computed analytically, while the finite integrals are computed numerically, as we will describe in the next section. The gamma matrices are chosen to anti-commute with $\gamma_5$, according to ref. [13]. This choice is not quite consistent with the axial anomaly [14]. But for the quenched calculation, as well as for non-singlet matrix elements, this should not matter. The results may differ though for the various schemes. Which scheme one will use finally will depend on the scheme that has been employed to compute the Wilson coefficients.

### Lattice Integrals

The lattice integrals are brought by the program into the form

$$\mathcal{B}(n_f, n_g; n_1, n_2, n_3, n_4) = a^d \int \frac{\mathrm{d}^d k}{(2\pi)^d} \frac{\hat{k}_1^{2n_1} \hat{k}_2^{2n_2} \hat{k}_3^{2n_3} \hat{k}_4^{2n_4}}{q^{n_f}(ak) g^{n_g}(ak)}, \tag{42}$$

where

$$\begin{aligned}
\hat{k}_\mu &= 2 \sin(ak_\mu/2), \\
\hat{k}^2 &= \sum_\mu \hat{k}_\mu^2, \\
q(ak) &= \sum_\mu \sin^2(ak_\mu) + (\hat{k}^2)^2, \\
g(ak) &= \hat{k}^2.
\end{aligned} \tag{43}$$

The integrals are infrared finite if

$$n_f + n_g < n_1 + n_2 + n_3 + n_4 + 2. \tag{44}$$

If they are divergent, we proceed as follows to extract the infrared poles. We notice that

$$\frac{1}{q(ak)} \stackrel{k \to 0}{=} \frac{1}{g(ak)} + O(1). \tag{45}$$

We then write

$$\mathcal{B}(n_f, n_g; n_1, n_2, n_3, n_4) = \mathcal{B}(0, n_f + n_g; n_1, n_2, n_3, n_4) \tag{46}$$
$$+ [\mathcal{B}(n_f, n_g; n_1, n_2, n_3, n_4) - \mathcal{B}(0, n_f + n_g; n_1, n_2, n_3, n_4)].$$

The first integral, $\mathcal{B}(0, n_f + n_g; n_1, n_2, n_3, n_4)$, which is purely gluonic, can be computed analytically [15], giving

$$\mathcal{B}(0, n; n_1, n_2, n_3, n_4) = \frac{b(n; n_1, n_2, n_3, n_4)}{\epsilon} + \widetilde{\mathcal{B}}(0, n; n_1, n_2, n_3, n_4), \tag{47}$$

where $b$ is some rational number divided by $\pi^2$, and $\widetilde{\mathcal{B}}$ can be reduced to a linear combination of one-dimensional elementary integrals, with exactly known coefficients, which can be computed to high precision.

The degree of infrared divergence of the remaining integral,

$$a^4 \int \frac{d^4 k}{(2\pi)^4} \hat{k}_1^{2n_1} \hat{k}_2^{2n_2} \hat{k}_3^{2n_3} \hat{k}_4^{2n_4} \left( \frac{1}{q^{n_f}(ak) g^{n_g}(ak)} - \frac{1}{g^{n_f + n_g}(ak)} \right), \tag{48}$$

is reduced by two. By repeating this procedure several times, all infrared divergences are shoveled into purely gluonic integrals, which again can be computed analytically. This leaves us with infrared finite fermionic integrals of the form (42) only.

As we already mentioned, these integrals will have to be computed numerically. Because the final expressions involve quite a few of these integrals, in particular for the higher operators, we need to know their values to a high degree of accuracy. The first method we tried was integration by Monte Carlo using VEGAS [16]. This led to large errors, in particular for the rank four operator. The second method we employed was four-dimensional Gauss-Legendre integration. Let us discuss and compare the two methods.

For definiteness, let us consider the integral

$$\mathcal{B}(2, 2; 3, 0, 0, 0) = a^4 \int_{-\pi/a}^{\pi/a} \frac{d^4 k}{(2\pi)^4} \frac{\hat{k}_1^6}{q^2(ak) g^2(ak)}. \tag{49}$$

Because the integrand is an even function of $k_\mu$, we may reduce the region of integration to the interval $[0, \pi/a]$. Using 50 iterations with approximately 80,000 integration points, VEGAS gives the result

$$\mathcal{B}(2, 2; 3, 0, 0, 0) = 0.0037450(5). \tag{50}$$

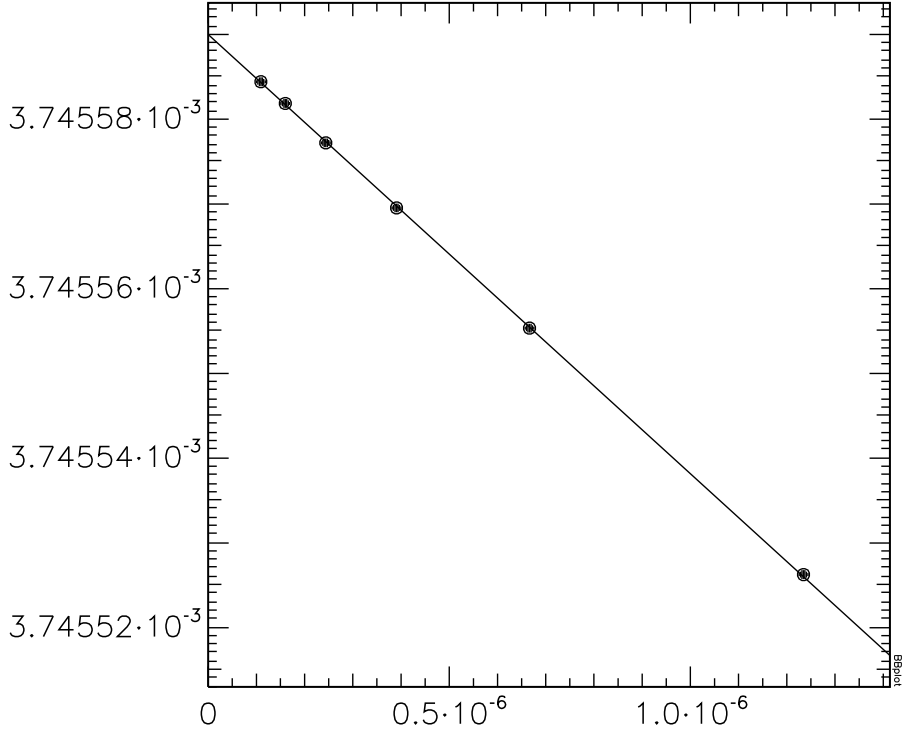

Figure 2: The lattice integral $\mathcal{B}(2,2;3,0,0,0)$ as a function of $1/N^4$. The line is a linear fit to the data points.

The Gauss-Legendre method distributes the integration points according to the zeroes of the Legendre polynomials. The density of integration points is particularly high near the endpoints. The endpoints themselves are, however, not covered. This feature is particularly well adapted to our type of integrals. We have computed the integral (49) using $N = 30, 35, 40, 45, 50$ and $55$ integration points in each direction. We expect that

$$\mathcal{B}[N] \cong \mathcal{B}[\infty] + \frac{\Delta \mathcal{B}}{N^4}. \tag{51}$$

The result is shown in fig. 2. The agreement with our expectation is very good. We find similarly good results for all other integrals as well. We may therefore fit eq. (51) to the data and take $\mathcal{B} = \mathcal{B}[\infty]$ as our final result, with the error being given by its variance. We then obtain

$$\mathcal{B}(2,2;3,0,0,0) = 0.0037455898(1). \tag{52}$$

This result is accurate to three more digits than the integration by Monte Carlo.

As an independent test we have computed the analytically known integral [15]

$$\mathcal{B}(0,1;1,1,0,0) \equiv Z_1 = 0.1077813135399. \quad (53)$$

With our method we obtain

$$Z_1 = 0.1077813135371(6), \quad (54)$$

which agrees with the result (53) up to the first 11 digits.

# 5  Results

In the quenched approximation and for the quark operators the renormalization constants can be written

$$Z_\mathcal{O}((a\mu)^2, g(a)) = 1 - \frac{g^2}{16\pi^2} C_F \left[ \gamma_\mathcal{O} \ln(a\mu) + B_\mathcal{O} \right], \quad C_F = \frac{4}{3}, \quad (55)$$

where $\gamma_\mathcal{O}$ is the anomalous dimension of the operator. The same anomalous dimensions must appear in the corresponding Wilson coefficients, so that the product of Wilson coefficient and operator matrix element is independent of $\mu$. Obviously, the anomalous dimensions do not depend on the particular choice of representation of $H(4)$ within a given $O(4)$ multiplet.

## Two Examples

Before we state our results, let us present two examples which may serve to illustrate in what aspects the lattice calculation differs from the continuum calculation.

*The First Moment*

Let us first consider the operator $\mathcal{O}_{\mu\nu}$. To order $g^2$ we obtain

$$\langle q(p) | \mathcal{O}_{\mu\nu} | q(p) \rangle = (1 + c_1) \gamma_\mu p_\nu + c_2 \gamma_\nu p_\mu + c_3 \gamma_\mu \delta_{\mu\nu} p_\nu + c_4 \not{p} \delta_{\mu\nu} + c_5 \frac{\not{p}}{p^2} p_\mu p_\nu - \text{traces}, \quad (56)$$

where all $c_i = O(g^2)$. See table 8 in the appendix for values of the coefficients. Note that the contribution with factor $c_3$ is non-$O(4)$ covariant. We shall look at two different representations:

$\underline{\mathcal{O}_{\{14\}}}$

This corresponds to the representation $\tau_3^{(6)}$. We obtain from eq. (56)

$$\langle q(p)|\mathcal{O}_{\{14\}}|q(p)\rangle = (1+c_1+c_2)\frac{1}{2}(\gamma_1 p_4 + \gamma_4 p_1) + c_5 \frac{\not{p}}{p^2} p_1 p_4. \tag{57}$$

The tree contribution, on the other hand, reads

$$\langle q(p)|\mathcal{O}_{\{14\}}|q(p)\rangle\Big|^{tree} = \frac{1}{2}(\gamma_1 p_4 + \gamma_4 p_1). \tag{58}$$

The standard renormalization procedure then leads to

$$\begin{aligned} Z\langle q(p)|\mathcal{O}_{\{14\}}|q(p)\rangle &= Z(1+c_1+c_2)\frac{1}{2}(\gamma_1 p_4 + \gamma_4 p_1) + Zc_5 \frac{\not{p}}{p^2} p_1 p_4 \\ &\equiv \frac{1}{2}(\gamma_1 p_4 + \gamma_4 p_1) + c_5 \frac{\not{p}}{p^2} p_1 p_4 + O(g^4). \end{aligned} \tag{59}$$

Making use of the fact that $Z = 1 + O(g^2)$, we obtain

$$Z = 1 - c_1 - c_2 + O(g^4). \tag{60}$$

There is no other operator transforming under $\tau_3^{(6)}$ with which this operator could mix.

$\underline{\mathcal{O}_{44} - \frac{1}{3}(\mathcal{O}_{11} + \mathcal{O}_{22} + \mathcal{O}_{33})}$

This corresponds to the representation $\tau_1^{(3)}$. Here we have

$$\begin{aligned} \langle q(p)|[\mathcal{O}_{44} - \frac{1}{3}(\mathcal{O}_{11} + \mathcal{O}_{22} + \mathcal{O}_{33})]|q(p)\rangle &= (1+c_1+c_2+c_3)[\gamma_4 p_4 - \frac{1}{3}(\gamma_1 p_1 + \gamma_2 p_2 \\ &+ \gamma_3 p_3)] + c_5 [p_4^2 - \frac{1}{3}(p_1^2 + p_2^2 + p_3^2)]\frac{\not{p}}{p^2}.(61) \end{aligned}$$

The tree contribution, on the other hand, reads

$$\langle q(p)|[\mathcal{O}_{44} - \frac{1}{3}(\mathcal{O}_{11} + \mathcal{O}_{22} + \mathcal{O}_{33})]|q(p)\rangle\Big|^{tree} = [\gamma_4 p_4 - \frac{1}{3}(\gamma_1 p_1 + \gamma_2 p_2 + \gamma_3 p_3)]. \tag{62}$$

The standard renormalization procedure then gives us

$$\begin{aligned} Z\langle q(p)|[\mathcal{O}_{44} - \frac{1}{3}(\mathcal{O}_{11} + \mathcal{O}_{22} + \mathcal{O}_{33})]|q(p)\rangle &= Z(1+c_1+c_2+c_3)[\gamma_4 p_4 - \frac{1}{3}(\gamma_1 p_1 + \gamma_2 p_2 \\ &+ \gamma_3 p_3)] + Zc_5 [p_4^2 - \frac{1}{3}(p_1^2 + p_2^2 + p_3^2)]\frac{\not{p}}{p^2} \\ &\equiv [\gamma_4 p_4 - \frac{1}{3}(\gamma_1 p_1 + \gamma_2 p_2 + \gamma_3 p_3)] \\ &+ c_5 [p_4^2 - \frac{1}{3}(p_1^2 + p_2^2 + p_3^2)]\frac{\not{p}}{p^2} + O(g^4). \end{aligned} \tag{63}$$

Making use of the fact that $Z = 1 + O(g^2)$, we obtain

$$Z = 1 - c_1 - c_2 - c_3 + O(g^4). \tag{64}$$

There is no other operator transforming under $\tau_1^{(3)}$ with which this operator could mix.

The origin of the difference of the renormalization constants (60), (64) is the presence of the non-$O(4)$ covariant contribution to (56).

*The Second Moment*

Let us now consider the operator $\mathcal{O}_{\mu\nu\rho}$. To order $g^2$ we obtain

$$\begin{aligned}\langle q(p)|\mathcal{O}_{\mu\nu\rho}|q(p)\rangle &= (1+c_1)\gamma_\mu p_\nu p_\rho + c_2\gamma_\nu p_\mu p_\rho + c_3\gamma_\rho p_\mu p_\nu + c_4\gamma_\mu \delta_{\nu\rho}p_\rho^2 + c_5\gamma_\nu \delta_{\mu\rho}p_\rho^2 \\ &+ c_6\gamma_\rho\delta_{\mu\nu}p_\nu^2 + c_7(\delta_{\mu\nu}\gamma_\mu p_\mu p_\rho + \delta_{\mu\rho}\gamma_\mu p_\mu p_\nu + \delta_{\nu\rho}\gamma_\nu p_\mu p_\nu) \quad (65)\\ &+ c_8\gamma_\mu\gamma_\nu\gamma_\rho(p_\nu^2 + p_\rho^2) + c_9\frac{\slashed{p}}{p^2}p_\mu p_\nu p_\rho + \cdots - \text{traces},\end{aligned}$$

where all $c_i = O(g^2)$. See table 9 in the appendix for values of the coefficients. The terms not listed here explicitly do not contribute to the specific operators which we will consider below. The contributions with factors $c_4 - c_8$ are non-$O(4)$ covariant.

$\underline{\mathcal{O}_{\{114\}} - \frac{1}{2}(\mathcal{O}_{\{224\}} + \mathcal{O}_{\{334\}})}$

This corresponds to the representation $\tau_1^{(8)}$. We obtain from eq. (65)

$$\begin{aligned}\langle q(p)|[\mathcal{O}_{\{114\}} - \frac{1}{2}(\mathcal{O}_{\{224\}} + \mathcal{O}_{\{334\}})]|q(p)\rangle &= (1+c_1+c_2+c_3)\frac{1}{3}[2\gamma_1 p_1 p_4 - \gamma_2 p_2 p_4 \\ &- \gamma_3 p_3 p_4 + \gamma_4 p_1^2 - \frac{1}{2}(\gamma_4 p_2^2 + \gamma_4 p_3^2)] \\ &+ (c_4+c_5+c_6+2c_8)\frac{1}{3}[\gamma_4 p_1^2 \\ &- \frac{1}{2}(\gamma_4 p_2^2 + \gamma_4 p_3^2)] \qquad (66)\\ &+ c_7[\gamma_1 p_1 p_4 - \frac{1}{2}(\gamma_2 p_2 p_4 + \gamma_3 p_3 p_4)] \\ &+ c_9\frac{\slashed{p}}{p^2}p_4[p_1^2 - \frac{1}{2}(p_2^2 + p_3^2)].\end{aligned}$$

The tree contribution, on the other hand, reads

$$\begin{aligned}\langle q(p)|[\mathcal{O}_{\{114\}} - \frac{1}{2}(\mathcal{O}_{\{224\}} + \mathcal{O}_{\{334\}})]|q(p)\rangle\Big|^{tree} &= \frac{1}{3}[2\gamma_1 p_1 p_4 - \gamma_2 p_2 p_4 - \gamma_3 p_3 p_4 \\ &+ \gamma_4 p_1^2 - \frac{1}{2}(\gamma_4 p_2^2 + \gamma_4 p_3^2)]. \qquad (67)\end{aligned}$$

This operator is not multiplicatively renormalizable. The non-$O(4)$ covariant contributions with factors $c_4 - c_8$ give rise to mixing with another operator of the same representation but of mixed symmetry.

$\underline{\mathcal{O}_{\langle\langle 411\rangle\rangle} - \frac{1}{2}(\mathcal{O}_{\langle\langle 422\rangle\rangle} + \mathcal{O}_{\langle\langle 433\rangle\rangle})}$

The operator $\mathcal{O}_{\{114\}} - \frac{1}{2}(\mathcal{O}_{\{224\}} + \mathcal{O}_{\{334\}})$ mixes with the operator $\mathcal{O}_{\langle\langle 411\rangle\rangle} - \frac{1}{2}(\mathcal{O}_{\langle\langle 422\rangle\rangle} + \mathcal{O}_{\langle\langle 433\rangle\rangle})$ which corresponds to the representation $\tau_1^{(8)}$ as well. ($\mathcal{O}_{\langle\langle\mu\nu\rho\rangle\rangle}$ was defined in eq. (29).) From eqs. (29), (65) we obtain

$$\langle q(p)|[\mathcal{O}_{\langle\langle 411\rangle\rangle} - \frac{1}{2}(\mathcal{O}_{\langle\langle 422\rangle\rangle} + \mathcal{O}_{\langle\langle 433\rangle\rangle})]|q(p)\rangle = [1+c_1-\frac{1}{2}(c_2+c_3)](-2\gamma_1 p_1 p_4 + \gamma_2 p_2 p_4$$

$$+\gamma_3 p_3 p_4 + 2\gamma_4 p_1^2 - \gamma_4 p_2^2 - \gamma_4 p_3^2)$$
$$+[c_4 - \frac{1}{2}(c_5 + c_6) + 2c_8][2\gamma_4 p_1^2 \qquad (68)$$
$$-\gamma_4 p_2^2 - \gamma_4 p_3^2].$$

The tree contribution of this operator reads

$$\langle q(p)|[\mathcal{O}_{\langle\langle 411\rangle\rangle} - \frac{1}{2}(\mathcal{O}_{\langle\langle 422\rangle\rangle} + \mathcal{O}_{\langle\langle 433\rangle\rangle})]|q(p)\rangle\Big|^{tree} = -2\gamma_1 p_1 p_4 + \gamma_2 p_2 p_4 + \gamma_3 p_3 p_4$$
$$+2\gamma_4 p_1^2 - \gamma_4 p_2^2 - \gamma_4 p_3^2. \qquad (69)$$

In an abbreviated form we express the renormalized operators as

$$\mathcal{O}_{\{\}}(\mu) = Z_{\{\}\{\}}\mathcal{O}_{\{\}}(a) + Z_{\{\}\langle\langle\rangle\rangle}\mathcal{O}_{\langle\langle\rangle\rangle}(a),$$
$$\mathcal{O}_{\langle\langle\rangle\rangle}(\mu) = Z_{\langle\langle\rangle\rangle\{\}}\mathcal{O}_{\{\}}(a) + Z_{\langle\langle\rangle\rangle\langle\langle\rangle\rangle}\mathcal{O}_{\langle\langle\rangle\rangle}(a). \qquad (70)$$

This is not a mixing in the usual sense as the matrix of anomalous dimensions is diagonal.

The renormalization conditions that follow from (66) – (69) are

$$\frac{1}{3} = Z_{\{\}\{\}}\frac{1}{3}(1 + c_1 + c_2 + c_3 + \frac{3}{2}c_7) - Z_{\{\}\langle\langle\rangle\rangle}[1 + c_1 - \frac{1}{2}(c_2 + c_3)]$$
$$= Z_{\{\}\{\}}\frac{1}{3} + \frac{1}{3}(c_1 + c_2 + c_3 + \frac{3}{2}c_7) - Z_{\{\}\langle\langle\rangle\rangle} + O(g^4), \qquad (71)$$
$$0 = Z_{\{\}\{\}}\frac{1}{3}(c_4 + c_5 + c_6 - \frac{3}{2}c_7 + 2c_8) + Z_{\{\}\langle\langle\rangle\rangle}[3 + 3c_1 - \frac{3}{2}(c_2 + c_3)$$
$$+2c_4 - c_5 - c_6 + 4c_8]$$
$$= \frac{1}{3}(c_4 + c_5 + c_6 - \frac{3}{2}c_7 + 2c_8) + 3Z_{\{\}\langle\langle\rangle\rangle} + O(g^4), \qquad (72)$$

where we have made use of the fact that $Z_{\{\}\langle\langle\rangle\rangle} = O(g^2)$ and $Z_{\{\}\{\}} = 1 + O(g^2)$. This then gives

$$Z_{\{\}\{\}} = 1 - c_1 - c_2 - c_3 - c_7 - \frac{1}{3}(c_4 + c_5 + c_6 + 2c_8) + O(g^4),$$
$$Z_{\{\}\langle\langle\rangle\rangle} = -\frac{1}{9}(c_4 + c_5 + c_6 - \frac{3}{2}c_7 + 2c_8) + O(g^4). \qquad (73)$$

Similarly, one finds for $Z_{\langle\langle\rangle\rangle\{\}}$ and $Z_{\langle\langle\rangle\rangle\langle\langle\rangle\rangle}$

$$Z_{\langle\langle\rangle\rangle\{\}} = -2c_4 + c_5 + c_6 - 4c_8 + O(g^4),$$
$$Z_{\langle\langle\rangle\rangle\langle\langle\rangle\rangle} = 1 - c_1 + \frac{1}{2}(c_2 + c_3) - \frac{1}{3}(2c_4 - c_5 - c_6 + 4c_8) + O(g^4). \qquad (74)$$

One can find other operators which do not give rise to mixing, but they require that the quark momentum is non-zero in more than one spatial direction.

## The operator $\mathcal{O}^5_{[2\{1]4\}}$

The operator $\mathcal{O}^5_{[2\{1]4\}}$ requires special attention. To one loop it does not mix. However it turns out that $\langle q(p)|\mathcal{O}^5_{[2\{1]4\}}|q(p)\rangle$ is linearly divergent, with the divergent part being given by

$$4.26568 \frac{g^2}{16\pi^2} C_F \frac{i}{a} \gamma_5 \gamma_2 (\gamma_4 p_1 - \gamma_1 p_4). \tag{75}$$

This contribution is directly proportional to $r$ and vanishes for naive fermions.

At first sight the occurrence of such a contribution might be surprising. But there is a good reason for that. In the OPE for $g_2$ we have suppressed one operator, namely [17]

$$\mathcal{O}^{5,m}_{[\sigma\{\mu_1]\cdots\mu_n\}} = m_q i \left(\frac{i}{2}\right)^n \bar{\psi}\gamma_{[\sigma}\gamma_{\{\mu_1]}\gamma_5 \overleftrightarrow{D}_{\mu_1} \cdots \overleftrightarrow{D}_{\mu_n\}} \psi - \text{traces}, \tag{76}$$

where $m_q$ is the quark mass. This explicitly quark mass dependent operator has also twist three. Usually, when the quark mass is renormalized multiplicatively, this operator can be neglected when one is only interested in the chiral limit. However, for Wilson fermions the situation is different. The divergent contribution (75) is exactly the contribution that is needed to renormalize the quark mass in the operator $\mathcal{O}^{5,m}_{[2\{1]4\}}$.

## Numerical Results

We shall now present our numerical results. We will first consider the operators listed in table 1. In the appendix we shall also give results for $\langle q(p)|\mathcal{O}_{\mu\nu}|q(p)\rangle$, $\langle q(p)|\mathcal{O}_{\mu\nu\rho}|q(p)\rangle$, $\langle q(p)|\mathcal{O}^5_{\mu\nu\rho}|q(p)\rangle$ and $\langle q(p)|\mathcal{O}_{\mu\nu\rho\sigma}|q(p)\rangle$ for general indices, from which the renormalization constants of all other representations [9] can be deduced.

The results for the anomalous dimensions $\gamma_\mathcal{O}$ are listed in table 2. The numbers agree with the anomalous dimensions known from the non-singlet Wilson coefficients [18].

The finite parts of the renormalization constants $B_\mathcal{O}$ are given in tables 3 – 6. Here we have also listed the individual contributions of the vertex, cockscomb, leg self-energy, leg tadpole and operator tadpole diagrams. (Remember that we are working in Feynman gauge.) The contribution of the leg tadpole diagram is

$$8\pi^2 \mathcal{B}(0,1;0,0,0,0) \equiv 8\pi^2 Z_0$$
$$= 12.23305015, \tag{77}$$

where we have used [15] $Z_0 = 0.1549333902311$. The total contribution of the operator tadpole diagrams is

$$- n_D 8\pi^2 Z_0, \tag{78}$$

where $n_D$ is the number of covariant derivatives of the operator. There are $n_D$ operator tadpole diagrams. In the case of the operators $\mathcal{O}_{\{114\}} - \frac{1}{2}(\mathcal{O}_{\{224\}} + \mathcal{O}_{\{334\}})$ and $\mathcal{O}_{\langle\langle 411\rangle\rangle} -$

| Operator | $\gamma_{\mathcal{O}}$ |
| :---: | :---: |
| $\mathcal{O}_{\{14\}}$ | $\frac{16}{3}$ |
| $\mathcal{O}_{\{44\}} - \frac{1}{3}(\mathcal{O}_{\{11\}} + \mathcal{O}_{\{22\}} + \mathcal{O}_{\{33\}})$ | $\frac{16}{3}$ |
| $\mathcal{O}_{\{114\}} - \frac{1}{2}(\mathcal{O}_{\{224\}} + \mathcal{O}_{\{334\}})$ | $\frac{25}{3}$ |
| $\mathcal{O}_{\langle\langle 411\rangle\rangle} - \frac{1}{2}(\mathcal{O}_{\langle\langle 422\rangle\rangle} + \mathcal{O}_{\langle\langle 433\rangle\rangle})$ | $\frac{7}{3}$ |
| $\mathcal{O}_{\{1144\}} + \mathcal{O}_{\{2233\}} - \mathcal{O}_{\{1133\}} - \mathcal{O}_{\{2244\}}$ | $\frac{157}{15}$ |
| $\mathcal{O}_2^5$ | $0$ |
| $\mathcal{O}_{\{214\}}^5$ | $\frac{25}{3}$ |
| $\mathcal{O}_{[2\{1]4\}}^5$ | $\frac{7}{3}$ |

Table 2: The operators and their anomalous dimensions.

$\frac{1}{2}(\mathcal{O}_{\langle\langle 422\rangle\rangle} + \mathcal{O}_{\langle\langle 433\rangle\rangle})$ the operator tadpole contribution is spread among $B_{\{\}\{\}}$ and $B_{\{\}\langle\langle\rangle\rangle}$. Here the result (78) holds for $c_1$, the order $g^2$ contribution with tree structure, and hence for $B_{\{\}\{\}} - 3B_{\{\}\langle\langle\rangle\rangle}$ and $B_{\langle\langle\rangle\rangle\langle\langle\rangle\rangle} - \frac{1}{3}B_{\langle\langle\rangle\rangle\{\}}$. The operator tadpole diagrams have the opposite sign to the leg tadpole diagrams. In the case of the operator $\mathcal{O}_{\mu\nu}$, which contains one covariant derivative, leg tadpole and operator tadpole diagrams cancel exactly. In all other cases the tadpole diagrams account for more than 60% of the total contribution.

The renormalization constants of the operators $\mathcal{O}_{\{14\}}$ and $\mathcal{O}_{\{44\}} - \frac{1}{3}(\mathcal{O}_{\{11\}} + \mathcal{O}_{\{22\}} + \mathcal{O}_{\{33\}})$ differ only in the vertex contribution to the finite part.

### Conversion to the $\overline{\text{MS}}$ Scheme

The Wilson coefficients are usually computed in the $\overline{MS}$ scheme, so that one would need to know the renormalization constants in this scheme too. The result in the $MS$ scheme is easily obtained. In table 7 we give the finite contribution of the continuum integrals to $B_{\mathcal{O}}$, which we denote by $B_{\mathcal{O}}^{con}$. Here $\gamma_E = 0.57721566$ is Euler's constant. The renormalization constants in the $MS$ scheme are then given by

$$B_{\mathcal{O}}^{MS} = B_{\mathcal{O}} - B_{\mathcal{O}}^{con}. \qquad (79)$$

Similarly, the renormalization constants in the $\overline{MS}$ scheme are obtained by

$$B_{\mathcal{O}}^{\overline{MS}} = B_{\mathcal{O}} - B_{\mathcal{O}}^{con} + \frac{\gamma_{\mathcal{O}}}{2}(\gamma_E - \ln 4\pi). \qquad (80)$$

| Diagram | $B_{\mathcal{O}_{\{14\}}}$ | $B_{\mathcal{O}_{\{44\}} - \frac{1}{3}(\mathcal{O}_{\{11\}} + \mathcal{O}_{\{22\}} + \mathcal{O}_{\{33\}})}$ |
|---|---|---|
| Vertex | 2.2930524(2) | 3.5753197(3) |
| Cockscomb | -5.0772671(1) | -5.0772671(1) |
| Leg Self-Energy | -0.3806456(7) | -0.3806456(7) |
| Leg Tadpole | $8\pi^2 Z_0$ | $8\pi^2 Z_0$ |
| Operator Tadpole | $-8\pi^2 Z_0$ | $-8\pi^2 Z_0$ |
| Total | -3.1648603(2) | -1.8825929(3) |

Table 3: The operators $\mathcal{O}_{\{14\}}$ and $\mathcal{O}_{\{44\}} - \frac{1}{3}(\mathcal{O}_{\{11\}} + \mathcal{O}_{\{22\}} + \mathcal{O}_{\{33\}})$.

| Diagram | $B_{\{\}\{\}}$ | $B_{\{\}\langle\langle\rangle\rangle}$ | $B_{\langle\langle\rangle\rangle\{\}}$ | $B_{\langle\langle\rangle\rangle\langle\langle\rangle\rangle}$ |
|---|---|---|---|---|
| Vertex | 1.357071(1) | -0.0559027(9) | -1.2849696(4) | 1.0080635(2) |
| Cockscomb | -6.73958572(2) | 0.21263441(7) | 7.4327048(3) | -2.0391253(2) |
| Leg Self-Energy | -0.3806456(7) | — | — | -0.3806456(7) |
| Leg Tadpole | $8\pi^2 Z_0$ | — | — | $8\pi^2 Z_0$ |
| Operator Tadpole | $\frac{2}{3}\pi^2 - \frac{64}{3}\pi^2 Z_0$ | $\frac{2}{9}\pi^2 - \frac{16}{9}\pi^2 Z_0$ | $4\pi^2 - 32\pi^2 Z_0$ | $\frac{4}{3}\pi^2 - \frac{80}{3}\pi^2 Z_0$ |
| Total | -19.571840(1) | -0.36847847(9) | -3.3060478(2) | -16.7960186(1) |

Table 4: The operators $\mathcal{O}_{\{114\}} - \frac{1}{2}(\mathcal{O}_{\{224\}} + \mathcal{O}_{\{334\}})$ and $\mathcal{O}_{\langle\langle 411\rangle\rangle} - \frac{1}{2}(\mathcal{O}_{\langle\langle 422\rangle\rangle} + \mathcal{O}_{\langle\langle 433\rangle\rangle})$.

| Diagram | $B_{\mathcal{O}_{\{1144\}} + \mathcal{O}_{\{2233\}} - \mathcal{O}_{\{1133\}} - \mathcal{O}_{\{2244\}}}$ |
|---|---|
| Vertex | -6.37131(2) |
| Cockscomb | -5.860427(1) |
| Leg Self-Energy | -0.3806456(7) |
| Leg Tadpole | $8\pi^2 Z_0$ |
| Operator Tadpole | $-24\pi^2 Z_0$ |
| Total | -37.07849(2) |

Table 5: The operator $\mathcal{O}_{\{1144\}} + \mathcal{O}_{\{2233\}} - \mathcal{O}_{\{1133\}} - \mathcal{O}_{\{2244\}}$.

| Diagram | $B_{\mathcal{O}_2^5}$ | $B_{\mathcal{O}_{\{214\}}^5}$ | $B_{\mathcal{O}_{[2\{1]4\}}^5}$ |
|---|---|---|---|
| Vertex | 3.94387868(6) | 0.661669(1) | 0.9887912(2) |
| Cockscomb | – | -7.6095645(2) | -4.0525425(2) |
| Leg Self-Energy | -0.3806456(7) | -0.3806456(7) | -0.3806456(7) |
| Leg Tadpole | $8\pi^2 Z_0$ | $8\pi^2 Z_0$ | $8\pi^2 Z_0$ |
| Operator Tadpole | – | $-16\pi^2 Z_0$ | $-16\pi^2 Z_0$ |
| Total | 15.79628324(6) | -19.561590(1) | -15.6774470(1) |

Table 6: The operators $\mathcal{O}_2^5$, $\mathcal{O}_{\{214\}}^5$ and $\mathcal{O}_{[2\{1]4\}}^5$.

| Operator | $B_{\mathcal{O}}^{con}$ |
|---|---|
| $\mathcal{O}_{\{14\}}$ | $-\frac{40}{9} + \frac{8}{3}(\gamma_E - \ln 4\pi)$ |
| $\mathcal{O}_{\{44\}} - \frac{1}{3}(\mathcal{O}_{\{11\}} + \mathcal{O}_{\{22\}} + \mathcal{O}_{\{33\}})$ | $-\frac{40}{9} + \frac{8}{3}(\gamma_E - \ln 4\pi)$ |
| $\mathcal{O}_{\{114\}} - \frac{1}{2}(\mathcal{O}_{\{224\}} + \mathcal{O}_{\{334\}})$ | $-\frac{67}{9} + \frac{25}{6}(\gamma_E - \ln 4\pi)$ |
| $\mathcal{O}_{\langle\langle 411\rangle\rangle} - \frac{1}{2}(\mathcal{O}_{\langle\langle 422\rangle\rangle} + \mathcal{O}_{\langle\langle 433\rangle\rangle})$ | $-\frac{35}{18} + \frac{7}{6}(\gamma_E - \ln 4\pi)$ |
| $\mathcal{O}_{\{1144\}} + \mathcal{O}_{\{2233\}} - \mathcal{O}_{\{1133\}} - \mathcal{O}_{\{2244\}}$ | $-\frac{2216}{225} + \frac{157}{30}(\gamma_E - \ln 4\pi)$ |
| $\mathcal{O}_2^5$ | 0 |
| $\mathcal{O}_{\{214\}}^5$ | $-\frac{67}{9} + \frac{25}{6}(\gamma_E - \ln 4\pi)$ |
| $\mathcal{O}_{[2\{1]4\}}^5$ | $-\frac{35}{18} + \frac{7}{6}(\gamma_E - \ln 4\pi)$ |

Table 7: The finite contribution of the continuum integrals to $B_{\mathcal{O}}$.

## 6  Summary

We have computed the renormalization constants of the leading twist lattice bilinear quark operators up to spin four. The calculation was done in the quenched approximation using Wilson fermions with $r = 1$. Results for other values of $r$ can be obtained from the authors.

For non-singlet quark operators and to one-loop order there is no difference between the quenched approximation and the full theory including dynamical quarks. The difference shows in the singlet operators, and here only in the renormalization constants $Z^{qg}$ and $Z^{gg}$ in eq. (31).

The renormalization constants for the axial vector current $\mathcal{O}_\mu^5$ and for the operator $\mathcal{O}_{\{14\}}$ were known before [19, 20]. The calculation of the renormalization constant for the operator $\mathcal{O}_{\{114\}} - \frac{1}{2}(\mathcal{O}_{\{224\}} + \mathcal{O}_{\{334\}})$ and its mixing parameters was done in parallel [21] to ours [2]. These authors use a slightly different basis of operators though. The results all agree.

We have explicitly stated the contributions that come from the tadpole diagrams. It is then straightforward to compute the renormalization constants in tadpole improved perturbation theory [3]. One possibility is to write

$$1 - \frac{g^2}{16\pi^2} C_F B_{\mathcal{O}} = \frac{u_F}{u_F^{n_D}} u_F^{n_D-1} (1 - \frac{g^2}{16\pi^2} C_F B_{\mathcal{O}}) = \frac{u_F}{u_F^{n_D}} (1 - \frac{g^{*\,2}}{16\pi^2} C_F \overline{B}_{\mathcal{O}}) + O(g^{*4}) \quad (81)$$

($n_D$: number of covariant derivatives), where

$$u_F = \frac{1}{3} \text{Tr} U_\mu = 1 - \frac{g^2}{16\pi^2} C_F 8\pi^2 Z_0 + O(g^4), \quad (82)$$

$U_\mu$ being the link matrix in Feynman gauge, and

$$\overline{B}_{\mathcal{O}} = B_{\mathcal{O}} + (n_D - 1) 8\pi^2 Z_0. \quad (83)$$

This reflects the fact that one finds $n_D$ operator tadpole and one leg tadpole diagrams, which are of the same magnitude but have opposite sign. In the case of the operators $\mathcal{O}_{\{114\}} - \frac{1}{2}(\mathcal{O}_{\{224\}} + \mathcal{O}_{\{334\}})$ and $\mathcal{O}_{\langle\langle 411\rangle\rangle} - \frac{1}{2}(\mathcal{O}_{\langle\langle 422\rangle\rangle} + \mathcal{O}_{\langle\langle 433\rangle\rangle})$, which mix, one has to consider $B_{\{\}\{\}} - 3 B_{\{\}\langle\rangle}$ and $B_{\langle\rangle\langle\rangle} - \frac{1}{3} B_{\langle\rangle\{\}}$ instead. One factor of $u_F$ will be absorbed into the normalization of the quark states, $\kappa_c \to \widetilde{\kappa}_c = \kappa_c u_F$, because $\widetilde{\kappa}_c$ is expected to have a better behaved perturbation series, $\widetilde{\kappa}_c = \frac{1}{8}(1 + \frac{g^{*\,2}}{4\pi} 0.066) + O(g^{*4})$. As the expansion parameter $g^*$ one uses the coupling constant renormalized at some physical scale. In ref. [5] we have compared our results with tadpole improved perturbation theory.

## Acknowledgment

We like to thank S. Capitani and G. Rossi for discussions. This work is supported in part by the Deutsche Forschungsgemeinschaft and the European Community under contract

number CHRX-CT92-0051.

# Appendix

In this appendix we shall present our results for the full tensors. We are only interested in the finite contributions. We write

$$\langle q(p)|\mathcal{O}_{\mu\nu}(a)|q(p)\rangle\Big|_{p^2=1/a^2} = \gamma_\mu p_\nu + \frac{g^2}{16\pi^2}C_F B_{\mu\nu} - \text{traces}, \tag{84}$$

$$\langle q(p)|\mathcal{O}_{\mu\nu\rho}(a)|q(p)\rangle\Big|_{p^2=1/a^2} = \gamma_\mu p_\nu p_\rho + \frac{g^2}{16\pi^2}C_F B_{\mu\nu\rho} - \text{traces}, \tag{85}$$

$$\langle q(p)|\mathcal{O}^5_{\mu\nu\rho}(a)|q(p)\rangle\Big|_{p^2=1/a^2} = \gamma_\mu \gamma_5 p_\nu p_\rho + \frac{g^2}{16\pi^2}C_F B^5_{\mu\nu\rho} - \text{traces}, \tag{86}$$

$$\langle q(p)|\mathcal{O}_{\mu\nu\rho\sigma}(a)|q(p)\rangle\Big|_{p^2=1/a^2} = \gamma_\mu p_\nu p_\rho p_\sigma + \frac{g^2}{16\pi^2}C_F B_{\mu\nu\rho\sigma} - \text{traces}, \tag{87}$$

where the logarithmic (anomalous) contributions are switched off, and

$$B_{\mu\nu} = \sum t_{\mu\nu}\, c, \tag{88}$$

$$B_{\mu\nu\rho} = \sum t_{\mu\nu\rho}\, c, \tag{89}$$

$$B^5_{\mu\nu\rho} = \sum t^5_{\mu\nu\rho}\, c^5, \tag{90}$$

$$B_{\mu\nu\rho\sigma} = \sum t_{\mu\nu\rho\sigma}\, c. \tag{91}$$

The numerical values of the coefficients $c, c^5$ for the finite contributions are listed in tables 8 – 11. We have omitted the singular contributions of order $1/a^3$ (in $B_{\mu\nu\rho\sigma}$), $1/a^2$ (in $B_{\mu\nu\rho\sigma}$, $B_{\mu\nu\rho}$ and $B^5_{\mu\nu\rho}$) and order $1/a$ (in $B_{\mu\nu\rho\sigma}$, $B_{\mu\nu\rho}$, $B^5_{\mu\nu\rho}$ and $B_{\mu\nu}$). They will cancel out after symmetrization and subtraction of the traces. The singular contributions may be obtained from the authors.

| $t_{\mu\nu}$ | $c$ |
|---|---|
| $\not{p}\,\delta_{\mu\nu}$ | 2.0523298(1) |
| $\delta_{\mu\nu}\,\gamma_\mu\,p_\mu$ | 1.2822674(5) |
| $\gamma_\nu\,p_\mu$ | -1.5824302(1) |
| $\gamma_\mu\,p_\nu$ | -1.5824302(1) |
| $\not{p}\,p_\mu\,p_\nu/p^2$ | -4/3 |

Table 8: The operator $\mathcal{O}_{\mu\nu}$.

| $t_{\mu\nu\rho}$ | $c$ | $t_{\mu\nu\rho}$ | $c$ |
|---|---|---|---|
| $\delta_{\nu\rho}\,\gamma_\mu\,p_\mu{}^2$ | 0.493942(1) | $\gamma_\mu\,p_\mu{}^2\,\delta_{\mu\nu\rho}$ | -0.00564(1) |
| $\delta_{\mu\rho}\,\gamma_\nu\,p_\mu{}^2$ | 0.138023(1) | $\delta_{\mu\nu}\,\gamma_\rho\,p_\mu{}^2$ | -0.886064(1) |
| $\not{p}\,\delta_{\mu\rho}\,p_\nu$ | 1.8345574(6) | $\not{p}\,\gamma_\mu\,\gamma_\rho\,p_\nu$ | -0.43037257(4) |
| $\delta_{\mu\rho}\,\gamma_\mu\,p_\mu\,p_\nu$ | 0.360814(2) | $\delta_{\nu\rho}\,\gamma_\nu\,p_\mu\,p_\nu$ | 0.360814(2) |
| $\gamma_\rho\,p_\mu\,p_\nu$ | -0.6741686(7) | $\delta_{\nu\rho}\,\gamma_\mu\,p_\nu{}^2$ | -3.051131(1) |
| $\delta_{\mu\rho}\,\gamma_\nu\,p_\nu{}^2$ | 0.493942(1) | $\gamma_\mu\,\gamma_\nu\,\gamma_\rho\,p_\nu{}^2$ | 0.5120432(1) |
| $\not{p}\,\delta_{\mu\nu}\,p_\rho$ | 0.9738123(5) | $\not{p}\,\gamma_\mu\,\gamma_\nu\,p_\rho$ | 0.43037257(4) |
| $\delta_{\mu\nu}\,\gamma_\mu\,p_\mu\,p_\rho$ | 0.360814(2) | $\gamma_\nu\,p_\mu\,p_\rho$ | -1.5349138(6) |
| $\gamma_\mu\,p_\nu\,p_\rho$ | -16.7985438(6) | $\delta_{\mu\nu}\,\gamma_\rho\,p_\rho{}^2$ | -0.5301448(9) |
| $\gamma_\mu\,\gamma_\nu\,\gamma_\rho\,p_\rho{}^2$ | 0.5120432(1) | $\not{p}\,p_\mu\,p_\nu\,p_\rho/p^2$ | -1 |
| $\delta_{\nu\rho}\,\gamma_\mu\,p^2$ | -0.3175361(2) | $\gamma_\mu\,\delta_{\mu\nu\rho}\,p^2$ | -0.0614063(8) |
| $\delta_{\mu\rho}\,\gamma_\nu\,p^2$ | 0.4422501(3) | $\delta_{\mu\nu}\,\gamma_\rho\,p^2$ | -0.2590425(2) |
| $\gamma_\mu\,\gamma_\nu\,\gamma_\rho\,p^2$ | 0.35064632(4) | $\not{p}\,\delta_{\nu\rho}\,p_\mu$ | 0.2185109(6) |
| $\not{p}\,p_\mu\,\delta_{\mu\nu\rho}$ | -0.494257(2) | | |

Table 9: The operator $\mathcal{O}_{\mu\nu\rho}$.

| $t^5_{\mu\nu\rho}$ | $c^5$ | $t^5_{\mu\nu\rho}$ | $c^5$ |
|---|---|---|---|
| $\not{p}\,\delta_{\nu\rho}\,\gamma_5\,p_\mu$ | -0.1618359(5) | $\not{p}\,\gamma_5\,p_\mu\,\delta_{\mu\nu\rho}$ | -0.329227(2) |
| $\delta_{\nu\rho}\,\gamma_\mu\,\gamma_5\,p_\mu{}^2$ | 0.064073(1) | $\gamma_\mu\,\gamma_5\,p_\mu{}^2\,\delta_{\mu\nu\rho}$ | 0.14207(1) |
| $\delta_{\mu\rho}\,\gamma_\nu\,\gamma_5\,p_\mu{}^2$ | 0.29828(1) | $\delta_{\mu\nu}\,\gamma_\rho\,\gamma_5\,p_\mu{}^2$ | -0.725806(1) |
| $\not{p}\,\delta_{\mu\rho}\,\gamma_5\,p_\nu$ | 1.644384(5) | $\not{p}\,\gamma_\mu\,\gamma_\rho\,\gamma_5\,p_\nu$ | -0.43037257(4) |
| $\delta_{\mu\rho}\,\gamma_\mu\,\gamma_5\,p_\mu\,p_\nu$ | 0.040298(2) | $\delta_{\nu\rho}\,\gamma_\nu\,\gamma_5\,p_\mu\,p_\nu$ | 0.525844(2) |
| $\gamma_\rho\,\gamma_5\,p_\mu\,p_\nu$ | -0.864342(6) | $\delta_{\nu\rho}\,\gamma_\mu\,\gamma_5\,p_\nu{}^2$ | -3.428208(1) |
| $\delta_{\mu\rho}\,\gamma_\nu\,\gamma_5\,p_\nu{}^2$ | 0.654199(1) | $\gamma_\mu\,\gamma_\nu\,\gamma_\rho\,\gamma_5\,p_\nu{}^2$ | 0.5120432(1) |
| $\not{p}\,\delta_{\mu\nu}\,\gamma_5\,p_\rho$ | 0.7836389(5) | $\not{p}\,\gamma_\mu\,\gamma_\nu\,\gamma_5\,p_\rho$ | 0.43037257(4) |
| $\delta_{\mu\nu}\,\gamma_\mu\,\gamma_5\,p_\mu\,p_\rho$ | 0.040298(2) | $\gamma_\nu\,\gamma_5\,p_\mu\,p_\rho$ | -1.7250872(6) |
| $\gamma_\mu\,\gamma_5\,p_\nu\,p_\rho$ | -16.9721616(5) | $\delta_{\mu\nu}\,\gamma_\rho\,\gamma_5\,p_\rho{}^2$ | -0.369887(9) |
| $\gamma_\mu\,\gamma_\nu\,\gamma_\rho\,\gamma_5\,p_\rho{}^2$ | 0.5120432(1) | $\not{p}\,\gamma_5\,p_\mu\,p_\nu\,p_\rho/p^2$ | -1 |
| $\delta_{\nu\rho}\,\gamma_\mu\,\gamma_5\,p^2$ | -0.1218283(2) | $\gamma_\mu\,\gamma_5\,\delta_{\mu\nu\rho}\,p^2$ | -0.160058(1) |
| $\delta_{\mu\rho}\,\gamma_\nu\,\gamma_5\,p^2$ | 0.6324235(3) | $\delta_{\mu\nu}\,\gamma_\rho\,\gamma_5\,p^2$ | -0.0688692(2) |
| $\gamma_\mu\,\gamma_\nu\,\gamma_\rho\,\gamma_5\,p^2$ | 0.35064632(4) | | |

Table 10: The operator $\mathcal{O}^5_{\mu\nu\rho}$.

| $t_{\mu\nu\rho\sigma}$ | $c$ | $t_{\mu\nu\rho\sigma}$ | $c$ |
|---:|---:|---:|---:|
| $\not{p}\,\delta_{\mu\rho}\,p_\sigma\,p_\nu$ | -3.08982(1) | $\not{p}\,p_\sigma\,p_\nu\,\delta_{\sigma\mu\rho}$ | 15.10485(4) |
| $\delta_{\nu\rho}\,\gamma_\sigma\,\gamma_\mu\,\not{p}^3$ | -0.2228411(4) | $\not{p}\,\delta_{\sigma\rho}\,\gamma_\sigma\,\gamma_\mu\,p_\sigma\,p_\nu$ | -1.4585045(5) |
| $\not{p}\,\delta_{\nu\rho}\,\gamma_\mu\,\gamma_\nu\,p_\sigma\,p_\nu$ | -1.4585045(5) | $\delta_{\mu\rho}\,\gamma_\sigma\,p_\sigma^2\,p_\nu$ | 6.77201(2) |
| $\gamma_\sigma\,p_\sigma^2\,p_\nu\,\delta_{\sigma\mu\rho}$ | -76.2628(2) | $\delta_{\sigma\rho}\,\gamma_\mu\,p_\sigma^2\,p_\nu$ | 7.5907(2) |
| $\delta_{\sigma\mu}\,\delta_{\nu\rho}\,\gamma_\nu\,p_\sigma^2\,p_\nu$ | -26.80781(6) | $\delta_{\sigma\mu}\,\gamma_\rho\,p_\sigma^2\,p_\nu$ | 7.35454(2) |
| $\delta_{\nu\rho}\,\gamma_\sigma\,\gamma_\mu\,\gamma_\nu\,p_\sigma^2\,p_\nu$ | 4.974603(2) | $\delta_{\sigma\rho}\,\gamma_\mu\,\gamma_\nu\,\not{p}^3$ | -0.2228411(4) |
| $\gamma_\sigma\,\gamma_\mu\,\gamma_\rho\,p_\sigma^2\,p_\nu$ | -0.4352333(4) | $\not{p}\,\delta_{\sigma\rho}\,p_\mu\,p_\nu$ | -3.84512(1) |
| $\delta_{\sigma\rho}\,\gamma_\sigma\,p_\sigma\,p_\mu\,p_\nu$ | 11.55906(4) | $\delta_{\mu\rho}\,\gamma_\mu\,p_\sigma\,p_\mu\,p_\nu$ | 15.55123(4) |
| $\delta_{\nu\rho}\,\gamma_\nu\,p_\sigma\,p_\mu\,p_\nu$ | 14.47606(4) | $\gamma_\rho\,p_\sigma\,p_\mu\,p_\nu$ | -4.66978(1) |
| $\delta_{\mu\rho}\,\gamma_\sigma\,p_\mu^2\,p_\nu$ | 5.45522(2) | $\delta_{\sigma\rho}\,\gamma_\mu\,p_\mu^2\,p_\nu$ | 7.90623(2) |
| $\not{p}\,\delta_{\sigma\mu}\,\delta_{\nu\rho}\,p_\nu^2$ | 7.73053(2) | $\not{p}\,\delta_{\nu\rho}\,\gamma_\sigma\,\gamma_\mu\,p_\nu^2$ | -0.699931(1) |
| $\not{p}\,\delta_{\sigma\rho}\,\gamma_\mu\,\gamma_\nu\,p_\nu^2$ | -0.5068522(3) | $\delta_{\sigma\mu}\,\delta_{\nu\rho}\,\gamma_\sigma\,p_\sigma\,p_\nu^2$ | -23.70894(6) |
| $\delta_{\nu\rho}\,\gamma_\mu\,p_\sigma\,p_\nu^2$ | 7.46065(2) | $\delta_{\mu\rho}\,\gamma_\nu\,p_\sigma\,p_\nu^2$ | 5.90631(2) |
| $\gamma_\nu\,p_\sigma\,p_\nu^2\,\delta_{\sigma\mu\rho}$ | -26.93646(6) | $\delta_{\sigma\rho}\,\gamma_\sigma\,\gamma_\mu\,\gamma_\nu\,p_\sigma\,p_\nu^2$ | 4.866452(2) |
| $\gamma_\mu\,\gamma_\nu\,\gamma_\rho\,p_\sigma\,p_\nu^2$ | -0.4352333(4) | $\delta_{\nu\rho}\,\gamma_\sigma\,p_\mu\,p_\nu^2$ | 4.9362(2) |
| $\delta_{\sigma\rho}\,\gamma_\nu\,p_\mu\,p_\nu^2$ | 6.92001(2) | $\delta_{\mu\rho}\,\gamma_\sigma\,p_\nu^3$ | 1.730699(6) |
| $\gamma_\sigma\,p_\nu^3\,\delta_{\sigma\mu\rho}$ | -7.89989(2) | $\delta_{\sigma\rho}\,\gamma_\mu\,p_\nu^3$ | 1.134569(7) |
| $\delta_{\sigma\mu}\,\delta_{\nu\rho}\,\gamma_\nu\,p_\nu^3$ | -28.7742(6) | $\delta_{\sigma\mu}\,\gamma_\rho\,p_\nu^3$ | 2.557857(6) |
| $\delta_{\nu\rho}\,\gamma_\sigma\,\gamma_\mu\,\gamma_\nu\,p_\nu^3$ | 3.764515(5) | $\gamma_\sigma\,\gamma_\mu\,\gamma_\rho\,p_\nu^3$ | -0.0321032(4) |
| $\not{p}\,\delta_{\mu\nu}\,p_\sigma\,p_\rho$ | -4.39407(1) | $\not{p}\,p_\sigma\,p_\rho\,\delta_{\sigma\mu\nu}$ | 15.11438(4) |
| $\not{p}\,\gamma_\mu\,\gamma_\nu\,p_\sigma\,p_\rho$ | 1.0825005(6) | $\delta_{\mu\nu}\,\gamma_\sigma\,p_\sigma^2\,p_\rho$ | 8.31511(2) |
| $\gamma_\sigma\,p_\sigma^2\,p_\rho\,\delta_{\sigma\mu\nu}$ | -76.258(2) | $\delta_{\sigma\nu}\,\gamma_\mu\,p_\sigma^2\,p_\rho$ | 6.3978(2) |
| $\delta_{\sigma\mu}\,\gamma_\nu\,p_\sigma^2\,p_\rho$ | 7.35454(2) | $\gamma_\sigma\,\gamma_\mu\,\gamma_\nu\,p_\sigma^2\,p_\rho$ | -1.317318(1) |
| $\not{p}\,\delta_{\sigma\nu}\,p_\mu\,p_\rho$ | -3.84512(1) | $\delta_{\sigma\nu}\,\gamma_\sigma\,p_\sigma\,p_\mu\,p_\rho$ | 11.54952(4) |
| $\delta_{\mu\nu}\,\gamma_\mu\,p_\sigma\,p_\mu\,p_\rho$ | 15.55123(4) | $\gamma_\nu\,p_\sigma\,p_\mu\,p_\rho$ | -5.53052(1) |
| $\delta_{\mu\nu}\,\gamma_\sigma\,p_\mu^2\,p_\rho$ | 7.35454(2) | $\delta_{\sigma\nu}\,\gamma_\mu\,p_\mu^2\,p_\rho$ | 7.90623(2) |
| $\not{p}\,\delta_{\sigma\mu}\,p_\nu\,p_\rho$ | -4.39407(1) | $\not{p}\,\gamma_\sigma\,\gamma_\mu\,p_\nu\,p_\rho$ | 1.0825005(6) |
| $\delta_{\sigma\mu}\,\gamma_\sigma\,p_\sigma\,p_\nu\,p_\rho$ | 15.55123(4) | $\gamma_\mu\,p_\sigma\,p_\nu\,p_\rho$ | -25.11451(1) |
| $\gamma_\sigma\,p_\mu\,p_\nu\,p_\rho$ | -3.36552(1) | $\delta_{\sigma\mu}\,\gamma_\nu\,p_\nu^2\,p_\rho$ | 8.31511(2) |
| $\gamma_\sigma\,\gamma_\mu\,\gamma_\nu\,p_\nu^2\,p_\rho$ | -1.206783(1) | $\delta_{\mu\nu}\,\gamma_\rho\,p_\sigma\,p_\rho^2$ | 8.31987(2) |
| $\gamma_\rho\,p_\sigma\,p_\rho^2\,\delta_{\sigma\mu\nu}$ | -26.95077(6) | $\gamma_\mu\,\gamma_\nu\,\gamma_\rho\,p_\sigma\,p_\rho^2$ | -0.259507(1) |
| $\delta_{\sigma\nu}\,\gamma_\rho\,p_\mu\,p_\rho^2$ | 6.92001(2) | $\delta_{\sigma\mu}\,\gamma_\rho\,p_\nu\,p_\rho^2$ | 8.31987(2) |
| $\gamma_\sigma\,\gamma_\mu\,\gamma_\rho\,p_\nu\,p_\rho^2$ | -0.259507(1) | $\delta_{\mu\nu}\,\gamma_\sigma\,p_\rho^3$ | 2.557857(6) |
| $\gamma_\sigma\,p_\rho^3\,\delta_{\sigma\mu\nu}$ | -8.12096(2) | $\delta_{\sigma\nu}\,\gamma_\mu\,p_\rho^3$ | 2.707691(7) |
| $\delta_{\sigma\mu}\,\gamma_\nu\,p_\rho^3$ | 2.557857(6) | $\gamma_\sigma\,\gamma_\mu\,\gamma_\nu\,p_\rho^3$ | -0.4456823(9) |
| $\not{p}\,p_\sigma\,p_\mu\,p_\nu\,p_\rho/p^2$ | -4/5 | $\not{p}\,\delta_{\sigma\rho}\,\delta_{\mu\nu}\,p^2$ | -2.656007(5) |

Continued on next page

| $t_{\mu\nu\rho\sigma}$ | $c$ | $t_{\mu\nu\rho\sigma}$ | $c$ |
|---|---|---|---|
| $\not{p}\,\delta_{\sigma\nu}\,\delta_{\mu\rho}\,p^2$ | -2.014185(6) | $\not{p}\,\delta_{\sigma\mu}\,\delta_{\nu\rho}\,p^2$ | -2.656007(5) |
| $\not{p}\,\delta_{\sigma\mu\nu\rho}\,p^2$ | 7.69008(2) | $\not{p}\,\delta_{\nu\rho}\,\gamma_\sigma\,\gamma_\mu\,p^2$ | 0.3209111(3) |
| $\not{p}\,\delta_{\sigma\rho}\,\gamma_\mu\,\gamma_\nu\,p^2$ | 0.3209111(3) | $\delta_{\sigma\rho}\,\delta_{\mu\nu}\,\gamma_\sigma\,p_\sigma\,p^2$ | 8.64832(2) |
| $\delta_{\sigma\nu}\,\delta_{\mu\rho}\,\gamma_\sigma\,p_\sigma\,p^2$ | 5.6632(2) | $\delta_{\sigma\mu}\,\delta_{\nu\rho}\,\gamma_\sigma\,p_\sigma\,p^2$ | 7.6552(2) |
| $\gamma_\sigma\,p_\sigma\,\delta_{\sigma\mu\nu\rho}\,p^2$ | -76.6905(2) | $\delta_{\nu\rho}\,\gamma_\mu\,p_\sigma\,p^2$ | -3.944894(7) |
| $\gamma_\mu\,p_\sigma\,\delta_{\sigma\nu\rho}\,p^2$ | 6.96961(2) | $\gamma_\mu\,p_\sigma\,\delta_{\mu\nu\rho}\,p^2$ | 7.67642(2) |
| $\delta_{\mu\rho}\,\gamma_\nu\,p_\sigma\,p^2$ | -1.351576(6) | $\gamma_\nu\,p_\sigma\,\delta_{\sigma\mu\rho}\,p^2$ | 7.72539(2) |
| $\delta_{\mu\nu}\,\gamma_\rho\,p_\sigma\,p^2$ | -2.635221(6) | $\gamma_\rho\,p_\sigma\,\delta_{\sigma\mu\nu}\,p^2$ | 7.72539(2) |
| $\delta_{\sigma\rho}\,\gamma_\sigma\,\gamma_\mu\,\gamma_\nu\,p_\sigma\,p^2$ | -1.49256(1) | $\gamma_\mu\,\gamma_\nu\,\gamma_\rho\,p_\sigma\,p^2$ | 0.4962342(4) |
| $\delta_{\nu\rho}\,\gamma_\sigma\,p_\mu\,p^2$ | -1.351576(6) | $\gamma_\sigma\,p_\mu\,\delta_{\sigma\nu\rho}\,p^2$ | 5.68442(2) |
| $\gamma_\sigma\,p_\mu\,\delta_{\mu\nu\rho}\,p^2$ | 7.72539(2) | $\delta_{\sigma\rho}\,\delta_{\mu\nu}\,\gamma_\mu\,p_\mu\,p^2$ | 7.6552(2) |
| $\delta_{\sigma\nu}\,\delta_{\mu\rho}\,\gamma_\mu\,p_\mu\,p^2$ | 7.87627(2) | $\delta_{\sigma\rho}\,\gamma_\nu\,p_\mu\,p^2$ | -1.993398(6) |
| $\delta_{\sigma\nu}\,\gamma_\rho\,p_\mu\,p^2$ | -1.993398(6) | $\delta_{\mu\rho}\,\gamma_\sigma\,p_\nu\,p^2$ | -2.344045(6) |
| $\gamma_\sigma\,p_\nu\,\delta_{\sigma\mu\rho}\,p^2$ | 7.67642(2) | $\delta_{\sigma\rho}\,\gamma_\mu\,p_\nu\,p^2$ | -2.312988(7) |
| $\delta_{\sigma\mu}\,\delta_{\nu\rho}\,\gamma_\nu\,p_\nu\,p^2$ | 8.64832(2) | $\delta_{\sigma\mu}\,\gamma_\rho\,p_\nu\,p^2$ | -2.635221(6) |
| $\delta_{\nu\rho}\,\gamma_\sigma\,\gamma_\mu\,\gamma_\nu\,p_\nu\,p^2$ | -1.49256(1) | $\gamma_\sigma\,\gamma_\mu\,\gamma_\rho\,p_\nu\,p^2$ | 0.4962342(4) |
| $\delta_{\mu\nu}\,\gamma_\sigma\,p_\rho\,p^2$ | -2.635221(6) | $\gamma_\sigma\,p_\rho\,\delta_{\sigma\mu\nu}\,p^2$ | 7.89749(2) |
| $\delta_{\sigma\nu}\,\gamma_\mu\,p_\rho\,p^2$ | -2.777849(7) | $\delta_{\sigma\mu}\,\gamma_\nu\,p_\rho\,p^2$ | -2.635221(6) |
| $\gamma_\sigma\,\gamma_\mu\,\gamma_\nu\,p_\rho\,p^2$ | 0.6418221(7) | $\not{p}\,\delta_{\sigma\rho}\,\delta_{\mu\nu}\,p_\sigma^2$ | 7.73053(2) |
| $\not{p}\,\delta_{\sigma\nu}\,\delta_{\mu\nu}\,p_\sigma^2$ | 6.33067(2) | $\not{p}\,\delta_{\sigma\mu}\,\delta_{\nu\rho}\,p_\sigma^2$ | 7.34437(2) |
| $\not{p}\,p_\sigma^2\,\delta_{\sigma\mu\nu\rho}$ | -73.0798(2) | $\not{p}\,\delta_{\nu\rho}\,\gamma_\sigma\,\gamma_\mu\,p_\sigma^2$ | -0.5044683(3) |
| $\not{p}\,\gamma_\sigma\,\gamma_\mu\,p_\sigma^2\,\delta_{\sigma\nu\rho}$ | -0.00238398(1) | $\not{p}\,\delta_{\sigma\rho}\,\gamma_\mu\,\gamma_\nu\,p_\sigma^2$ | -0.699931(1) |
| $\delta_{\sigma\rho}\,\delta_{\mu\nu}\,\gamma_\sigma\,p_\sigma^3$ | -28.7742(6) | $\delta_{\sigma\nu}\,\delta_{\mu\rho}\,\gamma_\sigma\,p_\sigma^3$ | -21.24517(6) |
| $\delta_{\sigma\mu}\,\delta_{\nu\rho}\,\gamma_\sigma\,p_\sigma^3$ | -25.46016(6) | $\gamma_\sigma\,p_\sigma^3\,\delta_{\sigma\mu\nu\rho}$ | 534.777(1) |
| $\delta_{\sigma\rho}\,\delta_{\mu\nu}\,\not{p}^3$ | 2.668754(6) | $\delta_{\nu\rho}\,\gamma_\mu\,p_\sigma^3$ | 2.653394(8) |
| $\gamma_\mu\,p_\sigma^3\,\delta_{\sigma\nu\rho}$ | -16.50226(8) | $\gamma_\mu\,p_\sigma^3\,\delta_{\mu\nu\rho}$ | -7.89989(2) |
| $\delta_{\mu\rho}\,\gamma_\nu\,p_\sigma^3$ | 1.666492(6) | $\gamma_\nu\,p_\sigma^3\,\delta_{\sigma\mu\rho}$ | -24.49405(7) |
| $\delta_{\mu\nu}\,\gamma_\rho\,p_\sigma^3$ | 2.557857(6) | $\gamma_\rho\,p_\sigma^3\,\delta_{\sigma\mu\nu}$ | -24.27298(7) |
| $\delta_{\sigma\rho}\,\gamma_\sigma\,\gamma_\mu\,\gamma_\nu\,p_\sigma^3$ | 3.875049(5) | $\gamma_\mu\,\gamma_\nu\,\gamma_\rho\,p_\sigma^3$ | -0.0321032(4) |
| $\delta_{\sigma\nu}\,\delta_{\mu\rho}\,\not{p}^3$ | 2.223071(6) | $\not{p}\,\delta_{\nu\rho}\,p_\sigma\,p_\mu$ | -3.84512(1) |
| $\not{p}\,p_\sigma\,p_\mu\,\delta_{\sigma\nu\rho}$ | 12.65199(4) | $\not{p}\,p_\sigma\,p_\mu\,\delta_{\mu\nu\rho}$ | 15.10485(4) |
| $\delta_{\sigma\mu}\,\delta_{\nu\rho}\,\not{p}^3$ | 2.668754(6) | $\delta_{\nu\rho}\,\gamma_\sigma\,p_\sigma^2\,p_\mu$ | 5.90631(2) |
| $\gamma_\sigma\,p_\sigma^2\,p_\mu\,\delta_{\sigma\nu\rho}$ | -63.6107(2) | $\gamma_\sigma\,p_\sigma^2\,p_\mu\,\delta_{\mu\nu\rho}$ | -27.148(6) |
| $\delta_{\sigma\rho}\,\delta_{\mu\nu}\,\gamma_\mu\,p_\sigma^2\,p_\mu$ | -23.71333(6) | $\delta_{\sigma\nu}\,\delta_{\mu\rho}\,\gamma_\mu\,p_\sigma^2\,p_\mu$ | -23.71809(6) |
| $\delta_{\sigma\rho}\,\gamma_\nu\,p_\sigma^2\,p_\mu$ | 6.33606(2) | $\delta_{\sigma\nu}\,\gamma_\rho\,p_\sigma^2\,p_\mu$ | 6.33606(2) |
| $\delta_{\mu\rho}\,\gamma_\sigma\,\gamma_\mu\,\gamma_\nu\,p_\sigma^2\,p_\mu$ | 0.2210686(2) | $\delta_{\sigma\mu\nu\rho}\,\not{p}^3$ | -8.90574(2) |

| $t_{\mu\nu\rho\sigma}$ | $c$ | $t_{\mu\nu\rho\sigma}$ | $c$ |
|---|---|---|---|
| $\not{p}\,\delta_{\sigma\rho}\,\delta_{\mu\nu}\,p_\mu{}^2$ | 7.34437(2) | $\not{p}\,\delta_{\sigma\nu}\,\delta_{\mu\rho}\,p_\mu{}^2$ | 6.33067(2) |
| $\delta_{\sigma\rho}\,\delta_{\mu\nu}\,\gamma_\sigma\,p_\sigma\,p_\mu{}^2$ | -26.80304(6) | $\delta_{\sigma\nu}\,\delta_{\mu\rho}\,\gamma_\sigma\,p_\sigma\,p_\mu{}^2$ | -17.05583(6) |
| $\delta_{\nu\rho}\,\gamma_\mu\,p_\sigma\,p_\mu{}^2$ | 7.90623(2) | $\gamma_\mu\,p_\sigma\,p_\mu{}^2\,\delta_{\sigma\nu\rho}$ | -23.85046(6) |
| $\gamma_\mu\,p_\sigma\,p_\mu{}^2\,\delta_{\mu\nu\rho}$ | -76.2672(2) | $\delta_{\mu\rho}\,\gamma_\nu\,p_\sigma\,p_\mu{}^2$ | 4.9362(2) |
| $\delta_{\mu\nu}\,\gamma_\rho\,p_\sigma\,p_\mu{}^2$ | 7.34977(2) | $\delta_{\nu\rho}\,\gamma_\sigma\,p_\mu{}^3$ | 1.666492(6) |
| $\gamma_\sigma\,p_\mu{}^3\,\delta_{\sigma\nu\rho}$ | -5.68185(2) | $\gamma_\sigma\,p_\mu{}^3\,\delta_{\mu\nu\rho}$ | -24.27298(7) |
| $\delta_{\sigma\rho}\,\delta_{\mu\nu}\,\gamma_\mu\,p_\mu{}^3$ | -25.46016(6) | $\delta_{\sigma\nu}\,\delta_{\mu\rho}\,\gamma_\mu\,p_\mu{}^3$ | -25.68122(6) |
| $\delta_{\sigma\rho}\,\gamma_\nu\,p_\mu{}^3$ | 2.112174(6) | $\delta_{\sigma\nu}\,\gamma_\rho\,p_\mu{}^3$ | 2.112174(6) |

Table 11: The operator $\mathcal{O}_{\mu\nu\rho\sigma}$. Here $\not{p}^3 = \sum_\mu \gamma_\mu p_\mu^3$.